\documentclass[12pt]{article}
\usepackage{amssymb,amsmath,amsbsy}
\usepackage{mathrsfs}
\usepackage{dsfont}
\usepackage{geometry}  
\usepackage{graphicx}
\usepackage{booktabs}
\usepackage{axodraw}
\usepackage{cite}
\usepackage{ulem}

\def\theequation{\arabic{section}.\arabic{equation}}

\newenvironment{Eqnarray}%
     {\arraycolsep 0.24em\begin{eqnarray}}{\end{eqnarray}}
\def\beq{\begin{equation}}
\def\eeq{\end{equation}}
\def\slashchar#1{\setbox0=\hbox{$#1$}           
   \dimen0=\wd0                                 
   \setbox1=\hbox{/} \dimen1=\wd1               
   \ifdim\dimen0>\dimen1                        
      \rlap{\hbox to \dimen0{\hfil/\hfil}}      
      #1                                        
   \else                                        
      \rlap{\hbox to \dimen1{\hfil$#1$\hfil}}   
      /                                         
   \fi}                                        %
\newcommand{\bea}{\begin{Eqnarray}}

\newcommand{\eea}{\end{Eqnarray}}
\newcommand{\ra}{\rightarrow}

\newcommand{\bra}[1]{\langle #1|}
\newcommand{\ket}[1]{|#1\rangle}

\newcommand{\lsim}{\raisebox{-0.13cm}{~\shortstack{$<$\\[-0.07cm] $\sim$}}~}
\newcommand{\gsim}{\raisebox{-0.13cm}{~\shortstack{$>$\\[-0.07cm] $\sim$}}~}
\def\ifmath#1{\relax\ifmmode #1\else $#1$\fi}

\def\ls#1{\ifmath{_{\lower1.5pt\hbox{$\scriptstyle #1$}}}}
\def\lsup#1{^{\lower 3pt\hbox{$\scriptstyle#1$}}}

\def\Eq#1{Eq.~(\ref{#1})}
\def\Eqs#1#2{Eqs.~(\ref{#1}) and (\ref{#2})}

\def\Rref#1{Ref.~\cite{#1}}

\def\Rrefs#1#2{Refs.~\cite{#1} and \cite{#2}}

\def\Fig#1{Fig.~\ref{#1}}
\def\sect#1{Section \ref{#1}}

\newcommand{\bsllp}{B_s^0 \rightarrow \ell^+ \ell'^{-}}

\newcommand{\cbsdllp}{\mathcal{B}(B_{s,d}^0 \rightarrow \ell^+ \ell'^{-})}

\newcommand{\bsdllp}{B_{s,d}^0 \rightarrow \ell^+ \ell'^{-}}

\newcommand{\bsdll}{B_{s,d}^0 \rightarrow \ell^+ \ell^{-}}
\newcommand{\bsdmumu}{B_{s,d}^0 \rightarrow \mu^+ \mu^{-}}
\newcommand{\bsmumu}{B_s^0 \rightarrow \mu^+ \mu^{-}}
\newcommand{\cbsmumu}{\mathcal{B}(B_s^0 \rightarrow \mu^+ \mu^{-})}
\newcommand{\cbdmumu}{\mathcal{B}(B_d^0 \rightarrow \mu^+ \mu^{-})}
\newcommand{\cbsdmumu}{\mathcal{B}(B_{s,d}^0 \rightarrow \mu^+ \mu^{-})}
\newcommand{\bs}{B_s^0 \rightarrow \mu^+ \mu^{-}}
\newcommand{\bd}{B_d^0 \rightarrow \mu^+ \mu^{-}}

\begin{document}
\tolerance=100000

\setcounter{page}{0}
\thispagestyle{empty}

\begin{flushright}
IPPP-08-96\\[-1mm]
arXiv:0812.4320\\[-1mm]
23 December  2008
\end{flushright}

\bigskip

\begin{center}
{\Large \bf Complete One-Loop MSSM Predictions for $B^0 \rightarrow
\ell^+ \ell'^{-}$ }\\[0.3cm] 
{\Large \bf at the Tevatron and LHC }\\[1.7cm]

{{\large Athanasios Dedes}$^{\, a,d}$, {\large Janusz Rosiek}$^{\,
b,d}$ {\large and Philip Tanedo}$^{\, c, d}$}\\[0.5cm]

{\it $^a$Division of Theoretical Physics, University of Ioannina,
Ioannina, GR 45110, Greece}\\[3mm]
{\it $^b$Institute of Theoretical Physics, University of Warsaw,
Ho{\.z}a 69, 00-681 Warsaw, Poland}\\[3mm]
{\it $^c$ Institute for High Energy Phenomenology, Newman Laboratory for Elementary Particle Physics, Cornell University, Ithaca, NY 14853, USA}\\[3mm]
{\it $^d$Institute for Particle Physics Phenomenology, University of
Durham, DH1 3LE, UK}

\end{center}

\vspace*{0.8cm}\centerline{\bf ABSTRACT}
\vspace{0.5cm}
\noindent{\small

During the last few years the Tevatron has dramatically improved the
bounds on rare $B$-meson decays into two leptons.  In the case of
$B_s^0 \rightarrow \mu^+ \mu^{-}$, the current bound is only ten times
greater than the Standard Model expectation.  Sensitivity to this
decay is one of the benchmark goals for LHCb performance and physics.
The Higgs penguin dominates this rate in the region of large
$\tan\beta$ of the MSSM. This is not necessarily the case in the
region of low $\tan\beta$, since box and $Z$-penguin diagrams may
contribute at a comparable rate.
In this article, we compute the complete one-loop MSSM contribution to
$\bsdllp$ for $\ell,\ell' = e, \mu$. We study the predictions for
general values of $\tan\beta$ with arbitrary flavour mixing parameters. 
We discuss the possibility of both enhancing and suppressing the branching 
ratios relative to their Standard Model expectations.
In particular, we find that there are ``cancellation regions'' in
parameter space where the branching ratio is  suppressed well below
the Standard Model expectation, making  it effectively invisible to
the LHC.  }

\vspace*{\fill}
\newpage

\setcounter{equation}{0}
\section{Introduction}
\label{sec:intro}

One of the most promising signals for new physics at the LHC is the
rare decay $B^0_{s} \rightarrow \mu^+\mu^-$.  This decay is suppressed
as a loop-level flavour-changing neutral current and by a lepton mass
insertion required for the final state muon helicities.  The LHC will
be the first experiment to be able to probe this decay channel all the
way down to its Standard Model (SM) branching ratio.
The decay is especially `clean' because its final state is easily
tagged and its only hadronic uncertainties come from the hadronic
decay constant $f_{B_{s}}$.
Further, enhancements to this branching ratio by new physics can be
resolved with only a few inverse femtobarns of data, making this an
exciting channel for beyond the standard model searches in the first
few years of LHC operation.

The current experimental status and the Standard Model predictions for
the branching ratios $\cbsdllp$ to leading order in QCD are displayed
in Table~\ref{tab1}. This is the updated version of the Table~1
presented in the review of \Rref{Dedes:2003kp}. Further reviews can be
found in \Rref{reviews}.

\begin{table*}[htb]\begin{center}\label{tab1}
\begin{tabular}{llll}\hline \hline
\textbf{Channel} & \textbf{Expt.} & \textbf{Bound (90\% CL)} & \textbf{SM Prediction} \\ \hline
$B^0_s \to \mu^+ \mu^-$ & CDF II~\cite{:2007kv}
                  & $<4.7\times 10^{-8}$
                  & $(4.8\pm 1.3) \times 10^{-9}$ \\ 
$B^0_d \to \mu^+ \mu^-$ & CDF II~\cite{:2007kv}
                  & $<1.5\times 10^{-8}$
                  & $(1.4\pm 0.4) \times 10^{-10}$  \\ 
\hline
$B^0_s \to \mu^+ e^-$ & CDF~\cite{Abe:1998bc}
                  & $<6.1\times 10^{-6}$
                  & $\approx 0$ \\ 
$B^0_d \to \mu^+ e^-$ & BABAR~\cite{Aubert:2007hb}
                  & $<9.2 \times 10^{-8}$
                  & $\approx 0$ \\
                         \hline 
\end{tabular}
\caption{Current experimental bounds and SM expectations for leptonic 
$B^0$-meson decays.}
\end{center}
\end{table*}
        
The Standard Model predictions for the dimuon decay of $B^0_s$ and
$B^0_d$ mesons were first calculated by Buchalla and Buras
in~\Rref{Buchalla:1993bv} and Higgs penguin contributions
in~\Rref{Bottela}.  Their analysis can be generalised to include
lepton flavour-violating decays with the final state $\mu^+e^-$ which
are not measurable within the SM extended with see-saw neutrino
masses.  The error in the SM predictions for the $B_{s,d}$ branching
ratios originates primarily from the uncertainties in the decay
constants~\cite{Gray:2005ad},
\begin{eqnarray}
f_{B_s}=230\pm 30 \hspace{.2cm} \mathrm{MeV} \;, \qquad
f_{B_d}=200\pm 30 \hspace{.2cm} \mathrm{MeV}\;,
\end{eqnarray}
and in the top-strange and top-down elements of the Cabibbo-Kobayashi-Maskawa (CKM)
matrix~\cite{Yao:2006px},
\begin{eqnarray}
|V_{ts}|=0.0406\pm 0.0027 \;, \qquad
|V_{td}|=0.0074\pm 0.0008\;.
\end{eqnarray}

The dimuon decay of $B^0_s$ is of particular interest to
experimentalists because it is a benchmark process for LHCb physics
and performance.  The LHCb will be able to directly probe the SM
predictions for this rare decay mode at $3\sigma$ ($5\sigma$)
significance with 2 $\mathrm{fb}^{-1}$ (6 $\mathrm{fb}^{-1}$) of data,
or after about one year (three years) of design
luminosity~\cite{arXiv:0710.5056}.  In addition, the general purpose
detectors ATLAS and CMS will also be able to reconstruct the $\bs$
signal with significance of $3\sigma$ after $\approx$ 30
$\mathrm{fb}^{-1}$~\cite{arXiv:0810.3618}. It is not clear whether LHC
can reach the SM expectation for $\bd$.

At the dawn of the LHC era, it is important to understand the possible
contributions of new physics to a discovery in the
$B^0_{s,d}\rightarrow \ell^+\ell'^-$ channels. These are particularly
promising decay channels for the Minimal Supersymmetric Standard Model
(MSSM).  Under the assumption of large values of $\tan\beta$ and
Minimal Flavour Violation (MFV), where the CKM matrix is the only
source of $CP$ and flavour violation, the branching ratio for $B^0_s
\rightarrow \mu^+\mu^-$ is dominated by the Higgs penguin mode and is
 approximately given by
\begin{eqnarray}
{\cal B}(B^0_s\rightarrow \mu^+\mu^-) &\approx& 5 \cdot 10^{-7}
\left(\frac{\tan\beta}{50}\right)^6\left(\frac{300 \hspace{.2cm}
\mathrm{GeV}}{M_A}\right)^4 \;,\label{eq:tanbeta}
\end{eqnarray}
where $M_A$ is the $CP$-odd Higgs mass. Thus in the large $\tan\beta$
regime this branching ratio can be significantly enhanced over the
Standard Model expectation.  This has been discussed extensively in
Refs.~\cite{Babu:1999hn, Bobeth:2001sq, Chankowski:2000ng,Huang,
Dedes:2003kp}.  The large $\tan\beta$ regime is preferred, for
example, by supersymmetric $SO(10)$ grand unified models.  Further,
the currently observed excess in the anomalous magnetic moment of the
muon $(g-2)_\mu$~\cite{Passera:2008hj} implies an additional
enhancement in $\cbsmumu$ in certain supergravity
scenarios~\cite{hep-ph/0108037}.  A field theoretic study of this
decay in the large $\tan \beta$ limit focusing on the resummation of
$\tan\beta$ was conducted in~\Rref{Dedes:2002er, Buras:2002vd,
Isidori:2001fv, Foster:2005kb}.

Thus far, however, the published analyses have focused primarily on
the large $\tan\beta$ region with MFV and have neglected possible
flavour mixing in the squark sector.
With the upcoming experimental probes of
$\mathcal{B}(B^0_{s,d}\rightarrow \ell^+\ell'^-)$ down to the SM
expectation, it is important to undertake a full, general calculation
of this branching ratio without \textit{a priori} assumptions on the
pattern of squark and slepton flavour mixing or electroweak symmetry
breaking.  In particular, in the region of \textit{low} $\tan
\beta$, the effects of box and $Z$-penguin diagrams could be of the
same order as the $\tan\beta$-enhanced Higgs penguins.  The
interference of these terms could conceivably lead to a cancellation
that would suppress the branching ratio \textit{below} the SM
prediction.  This region of parameter space has not yet been
thoroughly investigated. This paper fills the gap in the literature on
the low $\tan \beta$ properties of these decay modes.

If the branching ratio is significantly enhanced by new physics, it
may even be visible at the Tevatron. Alternately, if it is
significantly suppressed by new physics, it may be \textit{in}visible
even at the LHC. Either way, the status of this decay could become an
important factor for planned LHCb upgrades.  For example, it could
play a critical role in determining whether an LHCb upgrade should
focus on a more precise measurement of $\bs$ or instead reach for the
branching ratio of $\bd$ which is an order of magnitude smaller.

In this article we calculate MSSM predictions for dileptonic $B_d,B_s$
decays with arbitrary flavour mixing. In our numerical analysis, we
ignore $\tau$-lepton final states since decays like
$B_{s}^{0}\rightarrow {}\tau^{+}\tau^{-}$ or $B_{s}^{0}\rightarrow
\tau^{+}\mu^{-}$ since they cannot be observed accurately at the
Tevatron or LHC.  Although our calculation is sufficiently general to
include lepton flavour violating $B$-decays like $B^{0} \to \mu^\pm
e^\mp $, we do not consider them in our numerical analysis due to
their small branching ratio ($\lsim 10^{-11}$) at low
$\tan\beta$\footnote{Predictions of MSSM for $B \to \mu {}\, \tau$ or
$B \to \mu \, e$ at large $\tan\beta$ have been investigated in
\Rref{Dedes:2002rh}.}.  We therefore concentrate on the decays $\bs$
and $\bd$.  The general calculation is presented in the appendix and
the code used in our numerical analysis is available to the
public\footnote{In order to obtain the Fortran code, please send
e-mail to {\tt janusz.rosiek@fuw.edu.pl} }.

\setcounter{equation}{0}
\section{Effective Operators and Branching Ratios}
\label{sec:operators}

There are ten effective operators governing the dynamics of the
quarks-to-leptons transition $q^{I} q^{J} \rightarrow \ell^{+K}\,
\ell^{-L}$, with $q^{1}\equiv d, q^{2}\equiv s, q^{3}\equiv b$ and
$\ell^{1}\equiv e, \ell^{2}\equiv \mu, \ell^{3}\equiv \tau$.  The
effective Hamiltonian reads:
\begin{eqnarray}
{\cal H} \ = \ \frac{1}{(4\pi)^2}\sum_{X,Y=L,R}\biggl ( C_{VXY}\,
\mathcal{O}_{VXY} \ + \ C_{SXY} \, \mathcal{O}_{SXY} \ + \ C_{TX} \,
\mathcal{O}_{TX} \biggr ) \;, \label{ham}
\end{eqnarray}
where flavour and colour indices have been suppressed for brevity.
The (V)ector, (S)calar and (T)ensor operators are respectively given
by
\begin{eqnarray}
\mathcal{O}_{VXY}^{IJKL} \ &=& \ (\overline{q^{J}} \, \gamma^{\mu}\,
P_{X} \,q^{I} ) (\overline{\ell^{L}} \,\gamma_{\mu} \,P_{Y} \,
\ell^{K} ) \;, \nonumber \\[3mm]
\mathcal{O}_{SXY}^{IJKL} \ &=& \ (\overline{q^{J}} \, P_{X} \, q^{I} ) 
(\overline{\ell^{L}} \, P_{Y} \, \ell^{K} ) \;, \nonumber \\[3mm]
\mathcal{O}_{TX}^{IJKL} \ &=& \ (\overline{q^{J}} \sigma^{\mu\nu}\, 
P_{X}\, q^{I} ) (\overline{\ell^{L}}\, \sigma_{\mu\nu}\, \ell^{K} )
\;. \label{effops}
\end{eqnarray}
We follow the PDG conventions for the quark content of the
$B^0$-mesons, $B_{s}^{0}\equiv \overline{b}s$ and $B_{d}^{0}\equiv
\overline{b}d$ \cite{Amsler:2008zz}. Thus in~\Eq{effops} we identify
$q^{J}\equiv b$ and $q^{I} \equiv s$ or $d$ for $\bsdllp$
respectively.

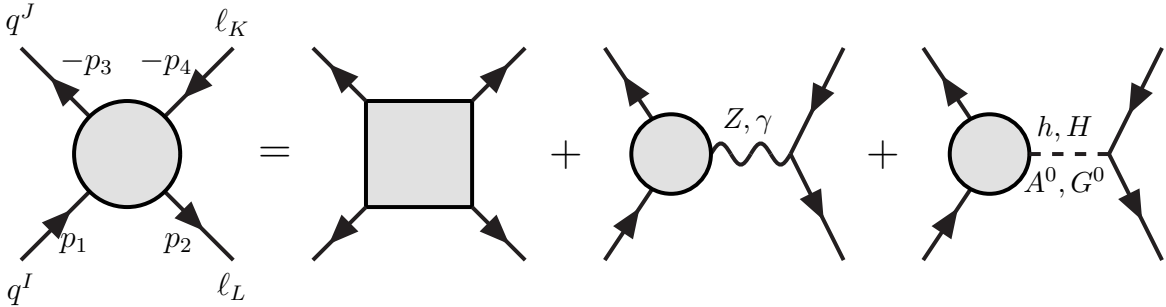
\begin{figure*}[htb]
\begin{center}
  \begin{picture}(445,112) (5,-14) \SetWidth{1.5}
    \ArrowLine(20,2)(50,32)
    \ArrowLine(70,32)(100,2)
    \ArrowLine(50,52)(20,82)
    \ArrowLine(100,82)(70,52)
    \GOval(60,42)(20,20)(0){0.882}
    \GBox(150,22)(190,62){0.882}
    \ArrowLine(150,62)(130,82)
    \ArrowLine(190,62)(210,82)
    \ArrowLine(190,22)(210,2)
    \ArrowLine(150,22)(130,2)
    \ArrowLine(260,52)(240,82)
    \ArrowLine(240,2)(260,32)
    \Photon(280,42)(310,42){4}{2}
    \ArrowLine(330,82)(310,42)
    \ArrowLine(310,42)(330,2)
    \GOval(265,42)(15,15)(0){0.882}
    \ArrowLine(380,52)(360,82)
    \ArrowLine(360,2)(380,32)
    \ArrowLine(450,82)(430,42)
    \ArrowLine(430,42)(450,2)
    \Text(110,42)[l]{\Large{{{{}}}{$=$}}}
    \Text(220,42)[l]{\Large{{{{}}}{$+$}}}
    \Text(340,42)[l]{\Large{{{{}}}{$+$}}}
    \Text(295,50)[b]{\small{{{{}}}{$Z,\gamma$}}}
    \GOval(385,42)(15,15)(0){0.882}
    \Text(20,87)[b]{\normalsize{{{{}}}{$q^J$}}}
    \Text(100,87)[b]{\normalsize{{{{}}}{$\ell_K$}}}
    \Text(20,-3)[t]{\normalsize{{{{}}}{$q^I$}}}
    \Text(100,-3)[t]{\normalsize{{{{}}}{$\ell_L$}}}
    \Text(35,72)[lb]{\normalsize{{{{}}}{$-p_3$}}}
    \Text(85,72)[rb]{\normalsize{{{{}}}{$-p_4$}}}
    \Text(35,12)[lt]{\normalsize{{{{}}}{$p_1$}}}
    \Text(85,12)[rt]{\normalsize{{{{}}}{$p_2$}}}
    \DashLine(400,42)(430,42){4}
    \Text(415,37)[t]{\small{{{{}}}{$A^0, G^0$}}}
    \Text(415,47)[b]{\small{{{{}}}{$h, H$}}}
  \end{picture}
\caption{Diagrams contributing to $q^I q^J\ra l^K l^L$ transitions.
\label{fig:diags}
}
\end{center}
\end{figure*}

The explicit forms of the Wilson coefficients for the MSSM are
calculated at the electroweak scale, $Q = M_{W}$. These are given in the
appendix.  The contributions to these coefficients can be classified
into $Z$-penguins, Higgs-penguins and box diagrams, shown in
Fig.~\ref{fig:diags}. The photon penguin contribution $\bsllp$
vanishes in matrix element calculations due to the Ward identity.  We
do not consider the very large $\tan\beta$ scenario ($\tan\beta\gsim 30$), 
since in this region our calculation has nothing to add to the current
literature (see previous section for references).  Thus no resummation
of higher orders in $\tan\beta$ is necessary and all formulae given in
the appendix are strictly one-loop.

%

We now focus on the decay $B_{s}^{0}\ra \ell^{+K}\,
\ell^{-L}$. Corresponding formulae for the $B_d^0$ decays can be
derived analogously.  Down quark vector and scalar currents hadronize
to $B_{s}$-mesons as
\begin{eqnarray}
\bra{0}\overline{b}\gamma_{\mu} P_{L(R)} s \ket{B_{s}(p)} \ &=& \ -(+) 
\frac{i}{2} p_{\mu} f_{B_{s}}
\;, \label{np1}  \\[3mm]
\bra{0}\overline{b} P_{L(R)} s \ket{B_{s}(p)} \ &=& \ +(-)
\frac{i}{2} \, \frac{M_{B_{s}}^{2} f_{B_{s}}}{m_{b}+m_{s}} \;.   \label{np2}
\end{eqnarray} 
where $p_{\mu}= p_{1\mu} + p_{3\mu}$ is the momentum of the decaying
$B_{s}$-meson of mass $M_{B_{s}}$.  In deriving~\Eq{np2} we have used
the quark equations of motion.  One immediate consequence of the
$\bsllp$ kinematics is that tensor operators vanish in the matrix
element $\bra{0}\overline{b}\,\sigma_{\mu\nu}\, s \ket{B_{s}(p)}$
because there is no way to make an antisymmetric tensor with the
single available momentum $p_{\mu}$.  This reduces the total number of
effective operators contributing to $\bsllp$ in~\Eq{effops} from ten
to eight.  The matrix element is therefore
\begin{eqnarray}\label{me}
\mathcal{M} \ =\ F_{S} \, \overline{\ell}\ell \ + \
        F_{P}\, \overline{\ell}\gamma_{5} \ell \ + \
        F_{V}\,  p^{\mu} \overline{\ell}\gamma_{\mu} \ell \ + \
        F_{A}\,  p^{\mu} \overline{\ell}\gamma_{\mu} \gamma_{5}\ell  \;,
\end{eqnarray}
where the $\ell$s correspond to external lepton spinors, e.g.  $F_{S}
\, \overline{\ell} \ell \equiv F_{S\, } \overline{u}(p_{2})\,
v(p_{4})$, etc.  The momenta are assigned as in \Fig{fig:diags}.
The (S)calar, (P)seudoscalar, (V)ector and (A)xial-vector form factors
in \Eq{me} are given by
\begin{eqnarray}
F_{S} \ &=& \ \frac{i}{4} \, \frac{M_{B_{s}}^{2}
f_{B_{s}}}{m_{b}+m_{s}} \; (C_{SLL} + C_{SLR} - C_{SRR} -C_{SRL} )
\;, \label{fs} \\[3mm]
F_{P} \ &=& \ \frac{i}{4} \, \frac{M_{B_{s}}^{2}
f_{B_{s}}}{m_{b}+m_{s}} \; (-C_{SLL} + C_{SLR} - C_{SRR} + C_{SRL} )
\;, \label{fp} \\[3mm]
F_{V} \ &=& \ -\frac{i}{4} f_{B_{s}} \; (C_{VLL} + C_{VLR} - C_{VRR} -
C_{VRL} ) \;, \label{fv} \\[3mm]
F_{A} \ &=& \ -\frac{i}{4} f_{B_{s}} \; (- C_{VLL} + C_{VLR} - C_{VRR}
+ C_{VRL} ) \;.   \label{fa}
\end{eqnarray}
It is now straightforward to square the matrix element in \Eq{me}, and
determine the branching ratio for the decay $\bsllp$,
\begin{eqnarray}
\mathcal{B}(B_s^0 \rightarrow \ell_L^- \, \ell_K^+) \ =\
\frac{\tau_{B_s}}{16\pi}\, \frac{|\mathcal{M}|^2}{M_{B_s}} \,
\sqrt{1-\biggl(\frac{m_{\ell_K} + m_{\ell_L}}{M_{B_s}}\biggr)^2 } \,
\sqrt{1-\biggl(\frac{m_{\ell_K} - m_{\ell_L}}{M_{B_s}}\biggr)^2 }
\;,\label{brr}
\end{eqnarray}
where $\tau_{B_s}$ is the lifetime of $B_s$ meson, and
\begin{eqnarray}
(4\pi)^4|\mathcal{M}|^2 &=& 2 |F_S|^2 \left [ M_{B_s}^2 - (m_{\ell_L} +
m_{\ell_K})^2 \right ] \ + \ 2 |F_P|^2 \left [ M_{B_s}^2 - (m_{\ell_L}
- m_{\ell_K})^2 \right ] \nonumber \\[3mm]
&+& 2 |F_V|^2 \left [M_{B_s}^2(m_{\ell_K} - m_{\ell_L})^2 -
(m_{\ell_K}^2 - m_{\ell_L}^2)^2 \right ] \nonumber \\[3mm]
&+& 2 |F_A|^2 \left [M_{B_s}^2(m_{\ell_K} + m_{\ell_L})^2 -
(m_{\ell_K}^2 - m_{\ell_L}^2)^2 \right ] \nonumber \\[3mm]
&+& 4\, \mathrm{Re} (F_S F_V^*) (m_{\ell_L} - m_{\ell_K}) \left
[M_{B_s}^2 + (m_{\ell_K} + m_{\ell_L})^2 \right ] \nonumber \\[3mm]
&+& 4\, \mathrm{Re} (F_P F_A^*) (m_{\ell_L} + m_{\ell_K})
\left [M_{B_s}^2 - (m_{\ell_L} - m_{\ell_K})^2 \right ] \;.   
\label{br}
\end{eqnarray} 

Notice that the contribution from the vector amplitude, $F_V$,
vanishes in the lepton flavour conserving case, $L=K$.  In this case
the formula in \Eq{br} agrees with results of~\Rref{Bobeth:2002ch}.

The form factors in Eqs.~(\ref{fs}\,--\,\ref{fa}) do not receive
additional renormalisation due to QCD corrections.  The conservation
of axial-vector current operators ($\mathcal{O}_{VXY}$) result in
vanishing anomalous dimension associated with this operator.  The
scalar operators ($\mathcal{O}_{SXY}$) renormalise like a quark mass
parameter and thus the ratio $C_{SXY}(Q)/[m_b(Q) + m_s(Q)]$ is a
renormalisation group invariant quantity~\cite{Hiller:2003js}.

Wilson coefficients and parameters entering
Eqs.~(\ref{fs}\,--\,\ref{fa}) and (\ref{br}) are all calculated at the
top quark mass scale, i.e. $Q=m_t$.  The quark pole masses, $m_b$ or
$m_t$, are related to their $\overline{DR}$-running one-loop quark
masses at the scale $Q$, $m_q(Q)$, by the well-known formulae:
\begin{eqnarray}
m_b(Q) \ &=& \ m_b\, \biggl [1 - \frac{5\,\alpha_s(m_b)}{3\pi} \biggr
]\, \biggl [\frac{\alpha_s(Q)}{\alpha_s(m_b)} \biggr ]^{\frac{4}{b_0}}
\;, \label{mb}\\[3mm]
m_t(m_t) \ &=& \ m_t\, \biggl [1 - \frac{5\, \alpha_s(m_t)}{3\pi}
\biggr ] \;, \label{mt}
\end{eqnarray} 
with $b_0 = 11 -2 n_f/3$ with $n_f =5$.  Since our calculation for the
SUSY corrections is performed in the $\overline{DR}$ renormalization
scheme~\cite{Capper:1979ns}, our initial conditions for parameters
must be converted into this scheme.  \Eqs{mb}{mt} contain the
appropriate $\overline{MS} \longrightarrow \overline{DR}$ conversion
factors~\cite{Martin:1993yx}. Similar conversions for gauge couplings
is small and is ignored in our numerical results.

We have included the general decays of~\Eq{brr} in our numerical code.
Due to small branching ratios, however, it is unlikely that we will
observe lepton flavour violating $B$-meson decays at moderate or small
values of $\tan\beta$ and so we will not consider these processes in
the remainder of this paper.

\setcounter{equation}{0}
\section{Numerical Analysis of $B_{s,d}\ra \mu^+\mu^-$}
\label{sec:analysis}

\subsection{Structure of the MSSM contributions}

We now focus on the lepton flavour-conserving processes $\bsdmumu$.
Recall from Section ~\ref{sec:intro} that we are interested in cases
where the branching ratios for these processes are either enhanced or
suppressed significantly relative to their Standard Model
predictions. An enhancement would either lead to an early discovery at
the LHC (or even the Tevatron) or stronger constraints on the allowed
magnitude of squark flavour violation. A suppression, on the other
hand, could lead to a non-observation of $\bsmumu$ at the LHCb due to
cancellations from new physics in the decay amplitude.

For the lepton flavour conserving decays $\ell_K=\ell_L=\mu$ and the
squared amplitude in \Eq{br} takes the form
\begin{eqnarray}
|\mathcal{M}|^2 \approx \frac{2 M_{B_q}^2}{(16\pi^2)^2} \left ( |F_S|^2 \ + \ |F_P + 2\,
m_\mu\, F_A |^2 \right ) \;,
\label{eq:ampsq}
\end{eqnarray}
where we have also taken the limit $m_\mu/M_{B_q} \to 0$.  We may
distinguish two possible scenarios for the relative size of the MSSM
contributions to the right-hand side of~\Eq{eq:ampsq}:
\vspace{1cm} 
\begin{itemize}
\item[1.] {\sl Higgs penguin domination or large $\tan\beta \gsim 10$}

In this large $\tan\beta$ regime one can usually expect an enhancement
of the branching ratios as in \Eq{eq:tanbeta}.  This case has been
thoroughly investigated in the literature, although mostly in the
limit of minimal flavour violation and vanishing intergenerational
squark mixing.  In such case it turns out that $|F_S| \approx |F_P|
\gg 2 m_\ell |F_A|$ because of $\tan^2\beta$ enhancements.  Although
this is the standard situation for large $\tan\beta$, it is not
general since a kind of Glashow -Iliopoulos -Maiani (GIM) cancellation 
mechanism may result in $F_{S,P}^{\rm SUSY} \approx 0$ ~\cite{Dedes:2002er, Foster:2005kb},
thus making the box and $Z$-penguin diagrams phenomenologically relevant.

\item[2.] {\sl Comparable Box, $Z$-penguin and Higgs penguin contributions or 
low $\tan\beta \lsim 10$}

In this low $\tan\beta$ case the supersymmetric Higgs-mediated form
factors $F_{S,P}$ are suppressed and become comparable to or even
smaller than $F_{A}$. Thus the full one-loop corrections to the
amplitude are needed. These are presented in the appendix. In this
case either an enhancement or a suppression of the branching ratios is
possible depending on the particular choice of MSSM parameters.
\end{itemize}

Barring accidental cancellations, an enhancement of the branching
ratios can come from any of the contributions in \Fig{fig:diags}.  On
the other hand, it is a bit trickier to suppress the branching ratios
below their Standard Model predictions as this requires a cancellation
between various terms. This is the case we would like to investigate
further.

We would like to find the minima of $\cbsdmumu$, i.e. the minima of 
\Eq{eq:ampsq}. We distinguish between two cases:
\begin{eqnarray}
&&F_{P} +  2\, m_\ell\, F_A \approx 0 \qquad  
\mathrm{and}\qquad  F_{P} \gg F_{S} \;, \label{c1}\\[2mm]
&&\mathrm{or}\nonumber\\[2mm] 
&&|F_S|\, \approx \, |F_P| \,\approx\, |F_A| \approx 0 \;.   \label{c2}
\end{eqnarray}

In the first case, \Eq{c1}, the pseudoscalar and axial contributions
cancel while the scalar contribution is negligible. This can be
realized, for example, in models where the MSSM is extended with an
additional, light, $CP$-odd Higgs boson. \Rref{Hodgkinson:2008qk}
shows that this can occur even in the minimal flavour violating limit
of such a model. Such cancellations, however, can also take place in
the general MSSM when left- and right-handed squarks mix in the
strange and charm sectors.
Furthrmore, it has been pointed out in \Rref{Alok:2008hh}, that
interference between the scalar/pseudoscalar new physics and Standard
Model operators can decrease the $\cbsmumu$ far below its SM
prediction.
This is explored further within MSSM in the numerical analysis of
\sect{sec:numset}.

The second case, \Eq{c2}, happens when Higgs contributions are
negligible compared to the axial contribution (i.e. low $\tan\beta$
and large $M_{A}$) and $F_{A}$ becomes small due to cancellations
among the $C_{VXY}$ coefficients in~\Eq{fv}.  Our numerical analysis
shows that such a cancellation is possible but requires a certain
amount of fine tuning once constraints on squark mass insertions from
other flavour-changing neutral current (FCNC) measurements are imposed.

\subsection{Numerical setup}
\label{sec:numset}

To quantitatively study the effects mentioned in the previous section,
we perform a scan over the MSSM parameter space.  The ranges of
variation over MSSM parameters are shown in Table~\ref{tab2}.  Because
our numerical analysis is based on the general calculation presented
in the previous section, we are not restricted to particular values of
$\tan\beta$ or the MFV scenario.
Flavour violation is parameterised by the ``mass insertions'', defined
as in \cite{Gabbiani:1996hi, Misiak:1997ei},
\begin{eqnarray}
\delta^{IJ}_{QXY} &=& \frac{(M^2_{Q})^{IJ}_{XY}}{\sqrt{
 (M^2_{Q})^{IJ}_{XX} (M^2_{Q})^{IJ}_{YY} }}\;. \label{eq:phys:massinsert}
\end{eqnarray}
As before, $I,J$ denote quark flavours, $X,Y$ denote superfield
chirality, and $Q$ indicates either the up or down quark superfield
sector.

\begin{table}[htb]
\begin{center}
\begin{tabular}{lcccc}
\hline \hline
Parameter & Symbol & Min & Max & Step \\ \hline
Ratio of Higgs vevs & $\tan\beta$ & 2 & 30 & varied \\
CKM phase & $\gamma$ & $0$& $\pi$ & $\pi/25$ \\
CP-odd Higgs mass & $M_{A}$ & 100 & 500 & 200 \\ 
SUSY Higgs mixing & $\mu$ & -450 & 450 & 300 \\ 
$SU(2)$ gaugino mass & $M_{2}$ & 100 & 500 & 200 \\ 
Gluino mass & $M_{3}$ &$3M_{2}$&$3M_{2}$& 0 \\ 
SUSY scale & $M_{\mathrm{SUSY}}$ & 500 & 1000 & 500 \\ 
Slepton Masses & $M_{\tilde{\ell}}$& $M_{\mathrm{SUSY}}/3$ &
$M_{\mathrm{SUSY}}/3$ & 0 \\
Left top squark mass & $M_{\tilde{Q}_{L}}$ & 200 & 500& 300 \\ 
Right bottom squark mass & $M_{\tilde{b}_{R}}$ & 200& 500& 300 \\
Right top squark mass & $M_{\tilde{t}_{R}}$ & 150 & 300 & 150 \\
Mass insertion & $\delta_{dLL}^{13}$, $\delta_{dLL}^{23}$ & -1& 1& 1/10 \\
Mass insertion & $\delta_{dLR}^{13}$, $\delta_{dLR}^{23}$ & -0.1& 0.1& 1/100 \\ \hline
\end{tabular} 
\caption{The range of input parameters for the numerical
scan.  ``SUSY scale'' refers to the common mass parameter for the
first two squark generations.  The parameter $\tan\beta$ takes on
values within the set: $\tan\beta = (2,4,6,8,10,13,16,19,22,25,30)$.
All mass parameters are in GeV. The top quark pole mass have been
taken 171 GeV. Imaginary part of parameters $\delta_{dLL}^{IJ}$,
$\delta_{dLR}^{IJ},\mu$ and $M_2$ have been set to zero. The
trilinear soft SUSY breaking couplings are set to
$A_{t}=A_{b}=M_{\tilde{Q}_{L}}$ and
$A_{\tilde{\tau}}=M_{\tilde{\ell}}$ throughout. }
\label{tab2}
\end{center}
\end{table}

To realistically estimate the allowed range for $\cbsdmumu$, one must
account the experimental constraints from measurements of many other
rare decays. SUSY mass insertions, in particular, are strongly
constrained by such measurements.
The most important constraints have been calculated in the framework
of the general MSSM using a standard set of
conventions~\cite{Misiak:1997ei, Pokorski:1999hz, Rosiek:1999mh,
Buras:2001mb, Buras:2002wq, Buras:2002vd, Buras:2004qb}. We have used
the library of numerical codes developed in those studies to bound the
MSSM parameter space based on the set of observables listed in
Table~\ref{tab3}; no further bounds (e.g. dark matter, electroweak 
observables, etc.) are imposed other than those listed.

For all the quantities in Table~\ref{tab3} for which the experimental
result and its error are known, we require
\bea
|Q^{exp} - Q^{th}| \leq 3\Delta Q^{exp} + q |Q^{th}|.
\label{eq:xacc}
\eea
For the quantities for which only the upper bound is known, we require
\bea
(1+q)|Q^{th}|\leq Q^{exp}.
\label{eq:xacc1}
\eea
%

\begin{table}[htb]
\begin{center}
\begin{tabular}{ccc}
\hline \hline
Quantity &  Current Measurement & Experimental Error \\ \hline
$m_{\chi^0_1}$ & $>$ 46 ~{\rm GeV} & \\
$m_{\chi^\pm_1}$ & $>$ 94 ~{\rm GeV} & \\
$m_{\tilde b}$ &  $>$ 89 ~{\rm GeV} & \\
$m_{\tilde t}$ & $>$ 95.7~{\rm GeV} & \\
$m_{h}$ & $>$ 92.8 ~{\rm GeV} & \\
$|\epsilon_{K}|$ & $2.232 \cdot 10^{-3}$ & $0.007 \cdot 10^{-3}$ \\
$|\Delta M_{K}|$ & $3.483 \cdot 10^{-15}$ & $0.006 \cdot 10^{-15}$ \\
$|\Delta M_{D}|$ & $ <0.46 \cdot 10^{-13}$ &  \\
$\Delta M_{B_d}$ & $3.337 \cdot 10^{-13}~{\rm GeV}$ & $0.033 \cdot
10^{-13}~{\rm GeV}$ \\
$\Delta M_{B_s}$ & $116.96 \cdot 10^{-13}~{\rm GeV}$ & $0.79 \cdot
10^{-13}~{\rm GeV}$ \\
Br($B\rightarrow X_{s} \gamma$) & $3.34\cdot 10^{-4}$ & $0.38 \cdot
10^{-4}$ \\
Br($K_{L}\rightarrow \pi^{0} \nu \bar{\nu}$) & $<1.5\cdot 10^{-10}$
&\\
Br($K^+\rightarrow \pi^+ \nu \bar{\nu}$) & $1.5\cdot 10^{-10}$ &
$1.3\cdot 10^{-10}$\\
Electron EDM & $<0.07\cdot 10^{-26}$ &\\
Neutron EDM & $<0.63\cdot 10^{-25}$ &\\
\hline \end{tabular}
\caption{Constraints used in the scan over MSSM parameters.   
LEP data are used for the Higgs mass bound~\cite{LEP}, i.e. $m_h\geq
92.8-114$ GeV depending on the value of $\sin^2(\alpha-\beta)$.}
\label{tab3} 
\end{center} 
\end{table}

The first and second terms on the right-hand side of \Eq{eq:xacc}
represent the $3\sigma$ experimental error and the theoretical error
respectively.  The latter differs from quantity to quantity and is
usually smaller than the value $q=50\%$ which we assume generically in
all calculations.
Apart from the theoretical errors that come from uncertainties in the
QCD evolution and hadronic matrix elements, one must also take into
account the limited density of a numerical scan.
In principle, with a very dense scan and sufficient computing time, it
should be possible to find SUSY parameters that fulfill \Eq{eq:xacc}
within the calculation's ``true'' theoretical errors.  This, however,
may not be necessary and may even be undesirable.  Our goal is to find
``generic'' values for the branching ratio $\cbsdmumu$, i.e.  values
allowed by fairly wide ranges of SUSY parameters without strong fine
tuning or the need to resort to special points in parameter space
where ``miraculous'' cancellations evade experimental bounds.  In our
scan we thus use wide ``theoretical'' errors assuming that this
procedure faithfully represents the ranges of the MSSM parameters. If
necessary the exact values of parameters fulfilling the bound in
\Eq{eq:xacc} with smaller $q$ can be found.  A more detailed
discussion of the problems associated with scanning over
multidimensional MSSM parameter space can be found
in~\cite{Buras:2004qb}.

\subsection{Predictions for ${\mathcal{B}}(B_s \to  \mu^+ \mu^-)$}

\begin{figure}[htb]
  \begin{center} \begin{tabular}{cc}
  \resizebox{75mm}{!}{\includegraphics{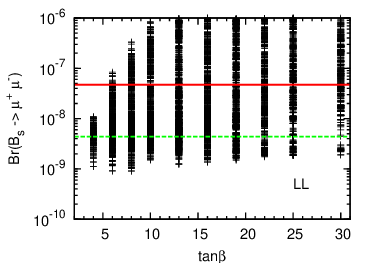}} &
  \resizebox{75mm}{!}{\includegraphics{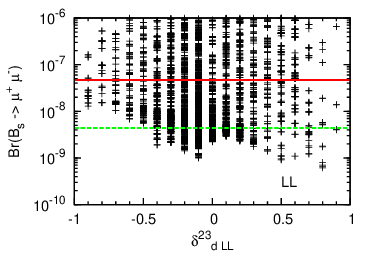}} \\
  \resizebox{75mm}{!}{\includegraphics{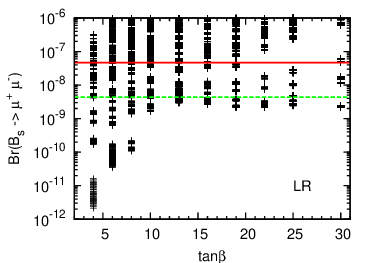}} &
  \resizebox{75mm}{!}{\includegraphics{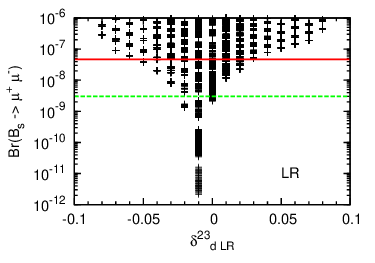}} \\
  \end{tabular}
\caption{ Upper panel: Predictions for ${\mathcal{B}}(B_s \to \mu^+ \mu^-)$ 
versus $\tan\beta$ (left) and $\delta_{d \, LL}^{23}$ (right) from the
scan of MSSM parameters in Table~\ref{tab2} and under the constraints
of Table~\ref{tab3}.  The upper solid line shows the current upper
bound from the Tevatron and the lower dashed line the SM expectation.  Lower
panel: Similar to the upper panel but with $\delta_{d \, LR}^{23}$
varied.}
    \label{fig2}
  \end{center}
\end{figure}

Fig.~\ref{fig2} shows the predictions for $\cbsmumu$ over a general
scan of 20 million points in parameter space according to
Table~\ref{tab3} and including the bounds described in the previous
section.  The upper bound set by CDF in Table~\ref{tab1}, depicted as
a solid red line, can be attained even with very low values of
$\tan\beta$.
We focus on the lower limit of the branching ratio and therefore
restrict to the region of parameter space where $\tan\beta \lsim 30$.
In this way we also avoid the technical complications connected with
the resummation of higher order terms, discussed in
Refs.~\cite{Dedes:2002er, Buras:2002vd, Isidori:2001fv,
Foster:2005kb}.  We vary $\delta_{d \, LL}^{23}$ (upper panel) and
$\delta_{d \, LR}^{23}$ (lower panel) one at a time while setting 
the other to zero,
e.g. all $\delta_{XY}^{{ij}}=0$ and only $\delta_{d \,
LL}^{23}\ne 0$ in the upper panel.

When $\delta_{d \, LL}^{23}$ is varied in the range $[-1, 1]$, we find
$\cbsmumu_{min} \approx 10^{-9}$.  This minimum is almost independent
of $\tan\beta$ but depends on the magnitude of the mass insertion
(upper right panel).  $|\delta_{d \, LL}^{23}|$ can take on values up
to $\approx 0.9$ and still pass all the constraints in
Table~\ref{tab3}, though points beyond $0.3$ are less dense. We note
here the importance of correctly incorporating the LEP Higgs mass
bound.  If for example we set $m_h > 114$ GeV independently of the
value of the $ZZH$ coupling, then $|\delta_{d \, LL}^{23}|$ is
restricted to values smaller than $\approx 0.3$.

More interesting is the case when $\delta_{d \, LR}^{23}$ is varied in
the range $[-0.1,0.1]$. We find a narrow cancellation region around
$\delta_{d \, LR}^{23}\approx-0.01$ and $\tan\beta \lsim 10$ where
$\cbsmumu_{min} \approx 10^{-12}$ (lower right panel). This is three
orders of magnitude lower than the Standard Model prediction, making
it effectively unobservable at the LHC.
In order to better understand cancellation region we study a
representative point with a very low branching ratio, for example:
\begin{eqnarray}
\tan\beta = 4, \quad 
M_{A} = 300, \quad \mu =-450, \quad M_{2}=100, \quad M_{3}=300 ,
\nonumber \\
\mathrm{SUSY ~scale}= 400, \quad M_{\tilde{t}_R}= 150, \quad 
A_{t,b}=M_{\tilde{t}_L}=
M_{\tilde{b}_{(L,R)}}=600 \;,\label{poi}
\end{eqnarray}
where all masses are in GeV.
\begin{figure}
  \begin{center}
    \begin{tabular}{cc}
      \resizebox{75mm}{!}{\includegraphics{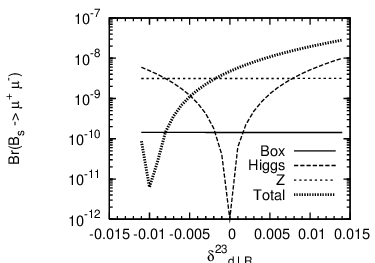}} &
      \resizebox{75mm}{!}{\includegraphics{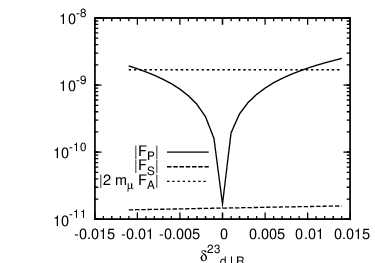}}
    \end{tabular}
\caption{Contributions to $\cbsdmumu$ from various parts with the
parameters in \Eq{poi}. Left: Contributions from the diagrams in
Fig.~\ref{fig:diags} versus $\delta^{23}_{d LR}$.  Right: Magnitude of the
form factors appearing in Eqs.~(\ref{fs}\,--\,\ref{fa}) versus $\delta^{23}_{d LR}$.}
\label{fig3}
\end{center}
\end{figure}
The cancellation is easy to understand if one independently considers
the contributions to the branching ratio from each diagram, as shown
on the left in Fig.~\ref{fig3}.
The `Box', `Higgs' and `$Z$' lines indicate the value of
$\cbsmumu$ given by only the listed contribution with all others set
to zero.  The total prediction for $\cbsmumu$ is also indicated.  We
observe that in the cancellation region
the Higgs- and $Z$-penguin magnitudes are comparable while the box
contribution is negligible.  This is suggestive of a cancellation
between the second and third class of diagrams in
Fig.~\ref{fig:diags}.
To observe this cancellation we individually plot the absolute
values of the form factors $F_{S,P}$ and $2 m_{\mu} F_{A}$ of
Eqs.~(\ref{fs}\,--\,\ref{fa}) in the right panel of Fig.~\ref{fig3}.
At the minimum point of the total branching ratio (thick-dashed
line in left panel of Fig.~\ref{fig3}) $|F_{P}|$ is approximately
equal to $|2 m_{\mu} F_{A}|$ and $|F_{S}|$ is negligibly small.  
This can be explained from the form of Eqs.~(\ref{fs}) and (\ref{fp}).
If one assumes $\delta^{23}_{d LR}=\left(\delta^{32}_{d
LR}\right)^\star$, then $C_{SLR}$ and $C_{SRL}$, the two Wilson
coefficients most sensitive to the variation of $\delta^{23}_{dLR}$,
have similar sizes and opposite sign and thus interfere destructively in
the amplitude.

Bounds on the $\delta$ parameters governing squark flavour mixing have
been presented in the literature using the mass insertion
approximation (MIA).  In
particular,~\Rrefs{Ciuchini:2006dx}{Ciuchini:2002uv} bound $|\delta_{d
\, LL}^{23}| \lsim 0.3$ and $|\delta_{d \, LR}^{23}| \lsim 0.02$ for a
particular point in the parameter space, $m_{\tilde{q}} = M_{3} = 350$
GeV. On the other hand, the results in Fig.~\ref{fig3} arise from an
extensive scan of the experimentally allowed parameter space without
resorting to MIA\footnote{Note that references to the
$\delta$-parameter in this paper are mainly for comparison and presentation.
Any other parameter that characterizes the squark mixing
would also be appropriate.  Recall that our calculation is not based on
expanding this parameter around zero and keeping
only leading terms (MIA approximation).  Instead, we numerically diagonalize all relevant
squark matrices and plug the result into the expressions given in the
Appendix.}. Thus the bounds on the $\delta$s presented here are both
different and more representative of the range of possibilities in the
general MSSM. The results of this scan show that $\delta^{23}_{dLL}$
is still rather weakly constrained, whereas $\delta^{23}_{d LR}\lsim
0.08$.

We remark here that varying $\delta^{13}_{dLL}$ or $\delta^{13}_{dLR}$
has almost no effect on $\cbsmumu$ which takes values along a narrow
band.

\subsection{Predictions for ${\mathcal{B}}(B_d \to  \mu^+ \mu^-)$}

\begin{figure}[htb]
  \begin{center} \begin{tabular}{cc}
  \resizebox{75mm}{!}{\includegraphics{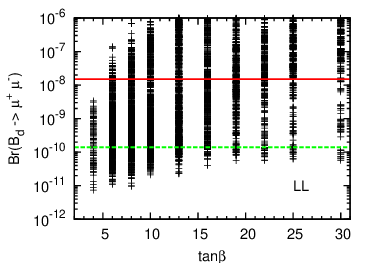}} &
  \resizebox{75mm}{!}{\includegraphics{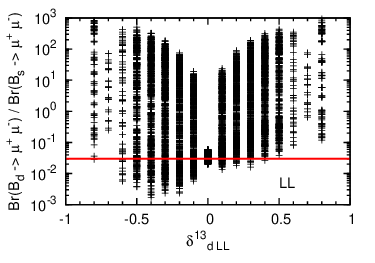}} \\
  \resizebox{75mm}{!}{\includegraphics{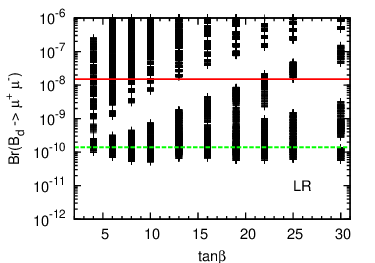}} &
  \resizebox{75mm}{!}{\includegraphics{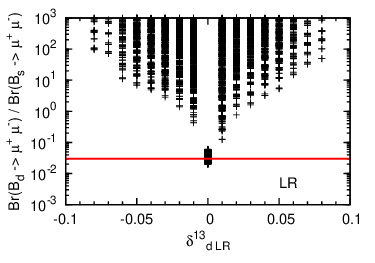}} \\
  \end{tabular}
\caption{ Upper panel: Predictions for
${\mathcal{B}}(B_d \to \mu^+ \mu^-)$ versus $\tan\beta$ (left) and
${\mathcal{B}}(B_d \to \mu^+ \mu^-)/{\mathcal{B}}(B_s \to \mu^+
\mu^-)$ versus $\delta_{d \, LL}^{13}$(right) from the scan of MSSM
parameters in Table~\ref{tab2} and under the constraints of
Table~\ref{tab3}.  The solid line shows the upper bound from the
Tevatron and the dashed line shows the SM expectation.  Lower panel:
Similar to the upper panel but with $\delta_{d \, LR}^{13}$ varied.
In all plots only the $\delta$ indicated is varied with all other 
mass insertions set to zero. }
\label{fig4} 
\end{center}
\end{figure}

We present the corresponding MSSM predictions for ${\mathcal{B}}(B_d
\to \mu^+ \mu^-)$  in Fig.~\ref{fig4} where $\delta_{d\, LL}^{13}$
or $\delta_{d\, LR}^{13}$ are varied instead of $\delta_{d\, LL}^{23}$
or $\delta_{d\, LR}^{23}$ along with the other SUSY parameters in
Table~\ref{tab2}.
Some sequences of points disappear due to the
experimental constraints given in Table~\ref{tab3}.  
%
Note that varying $\delta^{23}_{dLL}, \delta^{23}_{dLR}$ has almost no
effect on $\cbdmumu$.

For both cases there exist points where ${\mathcal{B}}(B_d \to \mu^+
\mu^-)$ is reduced by an order of magnitude relative to the SM. These
points are more sensitive to low $\tan\beta$ in the `LL' case and fall
into the case of \Eq{c2}.  It is also interesting to look at the the
ratio ${\mathcal{B}}(B_d \to \mu^+ \mu^-)/{\mathcal{B}}(B_s \to \mu^+
\mu^-)$ versus $\delta_{d\, LL}^{13}$ and $\delta_{d\, LR}^{13}$ ,
plotted in the right panel of Fig.~\ref{fig4}. Unlike the Standard
Model where ${\mathcal{B}}(B_d \to \mu^+ \mu^-)/{\mathcal{B}}(B_s \to
\mu^+ \mu^-) \approx |V_{td}/V_{ts}|^2 \leq 0.03$, the MSSM can
enhance this ratio by a factor of ten even for small values of
$\delta^{13}_{d\, LL}$ or $\delta^{13}_{d\, LR}$.  This suggests that
collider searches for ${\mathcal{B}}(B_d \to \mu^+ \mu^-)$ are as
important as those for ${\mathcal{B}}(B_s \to \mu^+ \mu^-)$. This
observation has been already discussed in the
literature~\cite{Bobeth:2002ch} in the leading $\tan\beta$
approximation. On the other hand MSSM can further reduce the ratio in
the `LL' case by an order of magnitude due to the aforementioned
cancellations in $\cbdmumu$.

\setcounter{equation}{0}
\section{Conclusions}
\label{sec:conclusion}

We have presented a complete, one-loop calculation of the branching
ratios for the rare decay modes $\bsdllp$ without resorting to the
limits of large $\tan\beta$, minimal flavour violation or SUSY
breaking scale dominance. Our final expressions are presented in an
appendix and are also available as a computer code (see footnote
2). We have used this code to perform a numerical exploration of the
MSSM parameter space for the modes $\bsdmumu$. We find that there
exist cancellation regions where the contribution of diagrams with
supersymmetric intermediate particles interferes destructively with
purely Standard Model diagrams, thus allowing the branching ratio to
be significantly smaller than the Standard Model prediction.  We
identify possible mechanisms of such cancellations and explain why
they can occur for certain regions of parameter space.  If
supersymmetry is a proper description of elementary interactions, such
effects may effectively hide the dimuon $B^0_s$ decay mode from the
LHCb even though it is supposed to be one of the experiment's
benchmark modes.  We have also shown that, barring the cancellations
mentioned above, supersymmetric contributions in the general MSSM
typically tend to enhance the branching ratio for $\bsdmumu$ even for
moderate values of $\tan\beta\lsim 10$ so that an experimental
measurement close to the SM prediction would put strong bounds on the
size of allowed flavour violation in the squark sector.  Finally, we
show that the $B^0_d\ra \mu^+\mu^-$ decay can also be either
suppressed or enhanced compared to its SM expectation, leading in some
cases to a situation where the rate of the $B^0_d$ decay is larger
then that of the $B^0_s$.

Our analysis provides a quantitative assessment of the viability of
non-minimal supersymmetric flavour structure and its consequences in
the neutral $B$-meson dimuon decay modes in light of existing
experimental constraints.  This is especially relevant due to recent
experimental hints for non-minimal flavour structure between the
second and third quark generations~\cite{Bona:2008jn,
Barberio:2008fa}.  Further, there have also been recent model-building
analyses of supersymmetric models not constrained to the `'Minimal
Flavour Violation'' scenario~\cite{Chankowski:2005jh, Nomura:2007ap,
Feng:2007ke}.

We conclude that new physics, in particular non-minimal flavour
violating supersymmetry, can manifest itself at future experiments as
either an enhancement or a suppression of the $\bsmumu$ decay rate
relative to the Standard Model.

\bigskip

\subsection*{Acknowledgements}

We would like to thank Frederic Teubert and Piotr Chankowski for
useful discussions, and Jennifer Girrbach for helpful comments on this manuscript.
A.D.  and J.R.  are partially supported by the RTN European Programme,
MRTN-CT-2006-035505 (HEPTOOLS, Tools and Precision Calculations for
Physics Discoveries at Colliders).
P.T. is supported by a Marshall Scholarship and a National Science
Foundation Graduate Research Fellowship.
J.R. was also supported in part by the Polish Ministry of Science and
Higher Education Grant No~1~P03B~108~30 for the years 2006-2008 and by
the EC 6th Framwework Programme MRTN-CT-2006-035863.
J.R. and P.T. would like to thank the Theoretical Physics Division at
the University of Ioannina for its generous hospitality.

\footnotesize

\newpage
\appendix
\renewcommand{\thesection}{Appendix~\Alph{section}}
\renewcommand{\theequation}{\Alph{section}.\arabic{equation}}

\setcounter{section}{0}
\setcounter{equation}{0}
\section{Wilson coefficients}
\label{appendix:wilson}

In this appendix we provide explicit results for the contributions to
the $\bsdll$ Wilson coefficients coming from the self-energy, Higgs-
and $Z$-penguin and box diagrams.  In general, Wilson coefficients
defined in \Eq{ham} can be decomposed as:
\bea
C_{VLL}^{IJKL} &=& B_{VLL}^{IJKL} - {e(1- 2s_W^2)\delta_{KL}\over 2
s_W c_W M_Z^2} \left( F_{ZL}^{IJ} - {e(1- \frac{2}{3}s_W^2)\over 2 s_W
c_W}\left(\Sigma_{dV}^{IJ} - \Sigma_{dA}^{IJ}\right) \right)\\[2mm]
C_{VRR}^{IJKL} &=& B_{VRR}^{IJKL} + {e s_W\delta_{KL}\over c_W M_Z^2}
\left( F_{ZR}^{IJ} + {e s_W\over 3 c_W}\left(\Sigma_{dV}^{IJ} +
\Sigma_{dA}^{IJ}\right) \right)\\[2mm]
C_{VLR}^{IJKL} &=& B_{VLR}^{IJKL} + {e s_W\delta_{KL}\over c_W M_Z^2}
\left( F_{ZL} - {e(1- \frac{2}{3}s_W^2)\over 2 s_W
c_W}\left(\Sigma_{dV}^{IJ} - \Sigma_{dA}^{IJ}\right) \right)\\[2mm]
C_{VRL}^{IJKL} &=& B_{VRL}^{IJKL} - {e(1- 2s_W^2)\delta_{KL}\over 2
s_W c_W M_Z^2} \left( F_{ZR}^{IJ} + {e s_W \over 3
c_W}\left(\Sigma_{dV}^{IJ} + \Sigma_{dA}^{IJ}\right) \right)\\[2mm]
C_{SLL}^{IJKL} &=& B_{SLL}^{IJKL} + {Y_l^K\delta_{KL}\over \sqrt{2}}
\sum_{k=1}^2 \left[{Z_R^{1k}\over m_{H_k^0}^2} \left( F_{HL}^{IJk} -
{Z_R^{1k}\over v_1} \left(\Sigma_{dS}^{IJ} - \Sigma_{dP}^{IJ}\right)
\right) \right.   \nonumber\\ 
&+&\left.    {Z_H^{1k}\over m_{A_k^0}^2} \left( F_{AL}^{IJk} +
{Z_H^{1k}\over v_1} \left(\Sigma_{dS}^{IJ} - \Sigma_{dP}^{IJ}\right)
\right) \right]\\[2mm]
C_{SRR}^{IJKL} &=& B_{SRR}^{IJKL} + {Y_l^K\delta_{KL}\over \sqrt{2}}
\sum_{k=1}^2 \left[{Z_R^{1k}\over m_{H_k^0}^2} \left( F_{HR}^{IJk} -
{Z_R^{1k}\over v_1} \left(\Sigma_{dS}^{IJ} + \Sigma_{dP}^{IJ}\right)
\right) \right.   \nonumber\\ 
&-&\left.    {Z_H^{1k}\over m_{A_k^0}^2} \left( F_{AR}^{IJk} -
{Z_H^{1k}\over v_1} \left(\Sigma_{dS}^{IJ} + \Sigma_{dP}^{IJ}\right)
\right) \right]\\[2mm]
C_{SLR}^{IJKL} &=& B_{SLR}^{IJKL} + {Y_l^K\delta_{KL}\over \sqrt{2}}
\sum_{k=1}^2 \left[{Z_R^{1k}\over m_{H_k^0}^2} \left( F_{HL}^{IJk} -
{Z_R^{1k}\over v_1} \left(\Sigma_{dS}^{IJ} - \Sigma_{dP}^{IJ}\right)
\right) \right.   \nonumber\\ 
&-&\left.    {Z_H^{1k}\over m_{A_k^0}^2} \left( F_{AL}^{IJk} +
{Z_H^{1k}\over v_1} \left(\Sigma_{dS}^{IJ} - \Sigma_{dP}^{IJ}\right)
\right) \right]\\[2mm]
C_{SRL}^{IJKL} &=& B_{SRL}^{IJKL} + {Y_l^K\delta_{KL}\over \sqrt{2}}
\sum_{k=1}^2 \left[{Z_R^{1k}\over m_{H_k^0}^2} \left( F_{HR}^{IJk} -
{Z_R^{1k}\over v_1} \left(\Sigma_{dS}^{IJ} + \Sigma_{dP}^{IJ}\right)
\right) \right.   \nonumber\\ 
&+&\left.    {Z_H^{1k}\over m_{A_k^0}^2} \left( F_{AR}^{IJk} -
{Z_H^{1k}\over v_1} \left(\Sigma_{dS}^{IJ} + \Sigma_{dP}^{IJ}\right)
\right) \right]\\[2mm]
C_{TL}^{IJKL} &=& B_{TL}^{IJKL}\\[2mm]
C_{TR}^{IJKL} &=& B_{TR}^{IJKL}
\eea
In the expression above $B_X$ are the box diagram contributions,
$F_{ZX}$ are the $Z$-penguin irreducible (triangle diagram)
contributions, $F_{HX}$ and $F_{AX}$ are respectively the
irreducible scalar and pseudoscalar Higgs penguin contributions and
$\Sigma_{dX}$ are the self energy contributions.
Indices are assigned as follows: $I,J$ are the
generation indices of quarks involved in the process, e.g. $I,J=(3,2)$ for
$B_s$ decay and $(3,1)$ for $B_d$ decay, and $K,L$ are the indices of
outgoing leptons, e.g.  $K,L=(2,2)$ for $B\ra \mu^+\mu^-$ decay etc.
For the definition of Higgs mixing matrices $Z_H, Z_R$, Higgs boson
masses and other symbols we refer reader to \Rref{Rosiek:1989rs},
the notation of which we use consistently in this appendix.

\subsection{Loop Integrals}

Here we collect the analytic forms of the relevant loop integrals for
this work.   
The two-point loop integral $B_1$ is defined as:
\begin{eqnarray}
\frac{1}{(4\pi)^2} p_\mu B_1(p,m^2,M^2) &=& \int \frac{d^dk}{(2\pi)^d} 
\frac{ik_\mu}{(k^2-m^2)\left[(k+p)^2 -  M^2\right]}.
\end{eqnarray}
The explicit formula for the 2-point loop integral $B_1$ at vanishing
external momentum is:
\begin{eqnarray}
 B_1(0,x,y) &=& \frac{1}{4} + \frac{1}{2}C_2(x,y,y),
\end{eqnarray}
where $C_2(x,y,y)$ is given in eq.~(\ref{eq:c2eq}).

The 3- and 4-point loop integrals at vanishing external momenta are
defined as:
\begin{eqnarray}
\frac{1}{(4\pi)^2} C_{2n}(m_1^2,m_2^2,m_3^2) &=& \int \frac{d^dk}{(2\pi)^d} 
\frac{ik^{2n}}{\prod_i^3(k^2-m_i^2)}\\
\frac{1}{(4\pi)^2} D_{2n}(m_1^2,m_2^2,m_3^2, m_4^2) &=& \int 
\frac{d^dk}{(2\pi)^d} \frac{ik^{2n}}{\prod_i^4(k^2-m_i^2)}.
\end{eqnarray}
The explicit formulae are listed below (we give also expressions for
some 3-point functions proportional to higher momenta powers, useful
in Higgs-penguin calculations):
\begin{eqnarray}
C_0(x,y,z) &=& \frac{y \log \frac yx }{(x-y)(z-y)} + \frac{z \log\frac
zx}{(x-z)(y-z)}\\
%
%
C_2(x,y,z) &=& \Delta + \log\frac{\mu^2}{x} + \frac{y^2 \log\frac
yx}{(x-y)(z-y)}+ \frac{z^2 \log\frac zx}{(x-z)(y-z)}\\
C_2(x,y,y) &=& \Delta + \log\frac{\mu^2}{y} + \frac{x}{x-y}\left[ 1 -
\frac{x\log \frac xy}{x-y} \right],
\label{eq:c2eq}
\\
C_{11}(x,y) &=& -{x - 3y\over 4(x-y)^2} + {y^2\over
2(x-y)^3}\log\frac{y}{x},\\
C_{12}(x,y) &=& -{x + y \over 2(x-y)^2} - {xy\over 3(x-y)^3}\log\frac{y}{x},
\end{eqnarray}
\begin{eqnarray}
D_0(x,y,z,t) &=& \frac{y \log \frac yx }{(y-x)(y-z)(y-t)} + \frac{z
\log\frac zx}{(z-x)(z-y)(z-t)}\nonumber \\
&& + \frac{t \log \frac tx }{(t-x)(t-y)(t-z)}\\
%
%
D_2(x,y,z,t) &=& \frac{y^2 \log \frac yx }{(y-x)(y-z)(y-t)} +
\frac{z^2 \log\frac zx}{(z-x)(z-y)(z-t)}\nonumber \\
&& + \frac{t^2 \log \frac tx }{(t-x)(t-y)(t-z)}
%
%
%
%
\end{eqnarray}
where the divergent piece $\Delta = \frac{2}{d-4} + \log(4\pi)\gamma_E
-1$ and $\mu$ is the renormalisation scale.

\subsection{Feynman Rules}
\label{sec:feyrul}

We use the following generic Feynman rules for the calculations below
($V_\mu$, $S_a$ and $f^j$ are generic vector bosons, scalars and
fermions, respectively):

	\begin{tabular}{ll}
	\begin{picture}(150,70)(0,-10)
	\DashArrowLine(10,0)(60,0){5}
	\Text(0,0)[c]{$S_a$}
	\ArrowLine(60,0)(110,0)
	\Text(120,0)[c]{$g^j$}
	\ArrowLine(60,50)(60,0)
	\Text(65,45)[l]{$f^i$}
	\Vertex(60,0){2}
	\end{picture}
	&
	\raisebox{15\unitlength}{
	\begin{minipage}{5cm}
	\lefteqn{
	i\left( V^{Liaj}_{fSg}P_L + V^{Riaj}_{fSg}P_R \right)
	}
	\end{minipage}
	}
	\end{tabular}

	\begin{tabular}{ll}
	\begin{picture}(150,70)(0,-10)
	\Photon(10,0)(60,0){3}{4}
	\Text(0,0)[c]{$V_{\mu}$}
	\DashArrowLine(60,0)(110,0){5}
	\Text(120,0)[c]{$S_b$}
	\Text(90,10)[c]{$k$}
	\DashArrowLine(60,50)(60,0){5}
	\Text(65,45)[l]{$R_a$}
	\Text(55,30)[r]{$p$}
	\Vertex(60,0){2}
	\end{picture}
	&
	\raisebox{35\unitlength}{
	\begin{minipage}{5cm}
	\lefteqn{
	iV^{ab}_{VRS} (p + k)^{\mu}
	}
	\end{minipage}
	}
	\end{tabular}
	
	\begin{tabular}{ll}
	\begin{picture}(150,70)(0,-10)
		\Photon(10,0)(60,0){3}{4}
		\Text(0,0)[c]{$V_{\mu}$}
		\ArrowLine(60,0)(110,0)
		\Text(120,0)[c]{$g^j$}
		\ArrowLine(60,50)(60,0)
		\Text(65,45)[l]{$f^i$}
		\Vertex(60,0){2}
	\end{picture}
	&
	\raisebox{35\unitlength}{
	\begin{minipage}{5cm}
		\lefteqn{
		i(V^{Lij}_{Vfg}\gamma_\mu P_L + V^{Rij}_{Vfg}\gamma_\mu P_R)
		}
	\end{minipage}
	}
	\end{tabular}	
	
	\begin{tabular}{ll}
	\begin{picture}(150,70)(0,-10)
	\DashArrowLine(10,0)(60,0){5}
	\Text(0,0)[c]{$Q_a$}
	\DashArrowLine(60,0)(110,0){5}
	\Text(120,0)[c]{$R_b$}
	\DashArrowLine(60,50)(60,0){5}
	\Text(65,45)[l]{$S_c$}
	\Vertex(60,0){2}
	\end{picture}
	&
	\raisebox{15\unitlength}{
	\begin{minipage}{5cm}
	\lefteqn{
	iV^{abc}_{QRS}
	}
	\end{minipage}
	}
	\end{tabular}

Explicit formulae for the generic couplings can be inserted
from~\cite{Rosiek:1989rs}.  In our calculations, $V$ can be $Z$ or $W$ bozon.
The indices $Q,R,S$ can denote $CP$-even or $CP-$odd Higgs bosons ($H^0_i,A^0_i$), squarks ($U_i,D_i$) or sleptons ($L_i, \tilde\nu^I$). 
The indices $f,g$ can denote quarks ($d^I,u^I$), leptons ($l^I,\nu^I$) or charginos and neutralinos ($C_i, N_i$).

\subsection{Box Diagram Contribution}
\label{appendix:box}

Box contributions to the Wilson coefficients are denoted by
$B_{ZXY}^{IJKL}$.  $Z$ labels the operator type, $Z = S,V,T$ for
scalar, vector or tensor, respectively.  $X$ and $Y$ label the handedness, $X,Y
\in \{L,R\}$.   $I,J$ and $K,L$ are quark and lepton generation indices,
as described at the beginning of the appendix.  Here and in the
following sections, we strictly follow the notation of~\cite{Rosiek:1989rs}, where expressions for all mixing matrices, vertices and other symbols used
can be found.

\bea
B_{VLL}^{IJKL} &=& {e^4\over 4 s_W^4} \sum_{M=1}^3 K^{MI} K^{MJ\star}
D_2(M^2_W,M^2_W,m^2_{u_M},0) \nonumber\\
&+& {e^2\over 4 s_W^2} \sum_{N=1}^3\sum_{m,n=1}^2\sum_{l=1}^6 Z_+^{1n}
Z_+^{1m\star} Z_{\tilde \nu}^{LN} Z_{\tilde \nu\star}^{KN}
V_{dUC}^{LIlm} V_{dUC}^{LJln\star}
D_2(m^2_{C_m},m^2_{C_n},m^2_{U_l},m^2_{{\tilde \nu}_N})\nonumber\\
&+&\frac{1}{4} \sum_{m,n=1}^4\sum_{l,o=1}^6 V_{dDN}^{LIlm}
V_{lLN}^{LKon} V_{dDN}^{LJln\star} V_{lLN}^{LLom\star}
D_2(m^2_{N_m},m^2_{N_n},m^2_{D_l},m^2_{L_o})\\
&+&\frac{1}{2} \sum_{m,n=1}^4\sum_{l,o=1}^6 V_{dDN}^{LIlm}
V_{lLN}^{LKom} V_{dDN}^{LJln\star} V_{lLN}^{LLon\star} m_{N_m} m_{N_n}
D_0(m^2_{N_m},m^2_{N_n},m^2_{D_l},m^2_{L_o})\nonumber
\eea
\bea
B_{VRR}^{IJKL} &=& \frac{1}{4} Y_l^K Y_l^L Y_d^I Y_d^J
\sum_{M=1}^3\sum_{k,l=1}^2 K^{MI} K^{MJ\star} (Z_H^{1k} Z_H^{1l})^2
D_0(m^2_{H^+_k},m^2_{H^+_l},m^2_{u_M},0) \nonumber\\
&+& \frac{1}{4} Y_l^K Y_l^L \sum_{N=1}^3\sum_{m,n=1}^2\sum_{l=1}^6
Z_-^{2m} Z_-^{2n\star} Z_{\tilde \nu}^{LN} Z_{\tilde \nu}^{KN\star}
V_{dUC}^{RIlm} V_{dUC}^{RJln\star}
D_2(m^2_{C_m},m^2_{C_n},m^2_{U_l},m^2_{{\tilde \nu}_N}) \nonumber\\
&+& \frac{1}{4} \sum_{m,n=1}^4\sum_{l,o=1}^6 V_{dDN}^{RIlm}
V_{lLN}^{RKon} V_{dDN}^{RJln\star} V_{lLN}^{RLom\star}
D_2(m^2_{N_m},m^2_{N_n},m^2_{D_l},m^2_{L_o}) \\
&+& \frac{1}{2} \sum_{m,n=1}^4\sum_{l,o=1}^6 V_{dDN}^{RIlm}
V_{lLN}^{RKom} V_{dDN}^{RJln\star} V_{lLN}^{RLon\star} m_{N_m} m_{N_n}
D_0(m^2_{N_m},m^2_{N_n},m^2_{D_l},m^2_{L_o})\nonumber
\eea
\bea
B_{VLR}^{IJKL} &=& \frac{1}{4} Y_l^K Y_l^L \sum_{M=1}^3\sum_{k,l=1}^2
(Y_u^M)^2 K^{MI} K^{MJ\star} Z_H^{1k} Z_H^{1l} Z_H^{2k} Z_H^{2l}
D_0(m^2_{H^+_k},m^2_{H^+_l},m^2_{u_M},0) \nonumber\\
&-& \frac{1}{2}Y_l^K Y_l^L \sum_{N=1}^3\sum_{m,n=1}^2\sum_{l=1}^6
Z_-^{2m} Z_-^{2n\star} Z_{\tilde \nu}^{LN} Z_{\tilde \nu}^{KN\star}
V_{dUC}^{LIlm} V_{dUC}^{LJln\star} m_{C_m} m_{C_n}
D_0(m^2_{C_m},m^2_{C_n},m^2_{U_l},m^2_{{\tilde \nu}_N})\nonumber\\
&-&\frac{1}{2} \sum_{m,n=1}^4\sum_{l,o=1}^6 V_{dDN}^{LIlm}
V_{lLN}^{RKon} V_{dDN}^{LJln\star} V_{lLN}^{RLom\star} m_{N_m} m_{N_n}
D_0(m^2_{N_m},m^2_{N_n},m^2_{D_l},m^2_{L_o}) \nonumber\\
&-&\frac{1}{4} \sum_{m,n=1}^4\sum_{l,o=1}^6 V_{dDN}^{LIlm}
V_{lLN}^{RKom} V_{dDN}^{LJln\star} V_{lLN}^{RLon\star}
D_2(m^2_{N_m},m^2_{N_n},m^2_{D_l},m^2_{L_o})
\eea
\bea
B_{VRL}^{IJKL} &=& -{e^2\over 2 s_W^2}
\sum_{N=1}^3\sum_{m,n=1}^2\sum_{l=1}^6 Z_+^{1n} Z_+^{1m\star}
Z_{\tilde \nu}^{LN} Z_{\tilde \nu}^{KN\star} V_{dUC}^{RIlm}
V_{dUC}^{RJln\star} m_{C_m} m_{C_n}
D_0(m^2_{C_m},m^2_{C_n},m^2_{U_l},m^2_{{\tilde \nu}_N})\nonumber\\
&-&\frac{1}{2} \sum_{m,n=1}^4\sum_{l,o=1}^6 V_{dDN}^{RIlm}
V_{lLN}^{LKon} V_{dDN}^{RJln\star} V_{lLN}^{LLom\star} m_{N_m} m_{N_n}
D_0(m^2_{N_m},m^2_{N_n},m^2_{D_l},m^2_{L_o})\nonumber\\
&-&\frac{1}{4} \sum_{m,n=1}^4\sum_{l,o=1}^6 V_{dDN}^{RIlm}
V_{lLN}^{LKom} V_{dDN}^{RJln\star} V_{lLN}^{LLon\star}
D_2(m^2_{N_m},m^2_{N_n},m^2_{D_l},m^2_{L_o})
\eea
\bea
B_{SLL}^{IJKL} &=& -{e\over 2s_W} Y_l^L \sum_{N=1}^3 \sum_{m,n=1}^2
\sum_{l=1}^6 Z_+^{1n} Z_-^{2m} Z_{\tilde \nu}^{LN} Z_{\tilde
\nu}^{KN\star} V_{dUC}^{LIlm} V_{dUC}^{RJln\star} m_{C_m} m_{C_n}
D_0(m^2_{C_m},m^2_{C_n},m^2_{U_l},m^2_{{\tilde \nu}_N})\nonumber\\
&-&\frac{1}{2} \sum_{m,n=1}^4\sum_{l,o=1}^6 (V_{lLN}^{LKon}
V_{lLN}^{RLom\star} +V_{lLN}^{LKom} V_{lLN}^{RLon\star})
V_{dDN}^{LIlm} V_{dDN}^{RJln\star} m_{N_m} m_{N_n}
D_0(m^2_{N_m},m^2_{N_n},m^2_{D_l},m^2_{L_o})\nonumber\\
\eea
\bea
B_{SRR}^{IJKL} &=& -{e\over 2 s_W} Y_l^K \sum_{N=1}^3 \sum_{m,n=1}^2
\sum_{l=1}^6 Z_+^{1m\star} Z_-^{2n\star} Z_{\tilde
\nu}^{KN\star} Z_{\tilde \nu}^{LN} V_{dUC}^{RIlm}
V_{dUC}^{LJln\star}m_{C_m} m_{C_n}
D_0(m^2_{C_m},m^2_{C_n},m^2_{U_l},m^2_{{\tilde \nu}_N})\nonumber \\
&-&\frac{1}{2} \sum_{m,n=1}^4\sum_{l,o=1}^6 (V_{lLN}^{RKon}
V_{lLN}^{LLom\star} +V_{lLN}^{RKom} V_{lLN}^{LLon\star})
V_{dDN}^{RIlm} V_{dDN}^{LJln\star} m_{N_m} m_{N_n}
D_0(m^2_{N_m},m^2_{N_n},m^2_{D_l},m^2_{L_o})\nonumber\\
\eea
\begin{eqnarray}
B_{SLR}^{IJKL} &=& -{e^2\over 2s_W^2} Y_l^K Y_d^J \sum_{M=1}^3
\sum_{n=1}^2 K^{MI} K^{MJ\star} (Z_H^{1n})^2
D_2(m^2_{u_M},m^2_{H^+_n},M^2_W,0) \\
&-& {e\over 2s_W} Y_l^K \sum_{N=1}^3 \sum_{m,n=1}^2
\sum_{l=1}^6 Z_+^{1m\star} Z_-^{2n\star} Z_{\tilde
\nu}^{LN} Z_{\tilde \nu}^{KN\star} V_{dUC}^{LIlm} V_{dUC}^{RJln\star} 
D_2(m^2_{C_m},m^2_{C_n},m^2_{U_l},m^2_{{\tilde \nu}_N})\nonumber\\
&-&\frac{1}{2} \sum_{m,n=1}^4\sum_{l,o=1}^6 (V_{lLN}^{RKon}
V_{lLN}^{LLom\star} +V_{lLN}^{RKom} V_{lLN}^{LLon\star})
V_{dDN}^{LIlm} V_{dDN}^{RJln\star}
D_2(m^2_{N_m},m^2_{N_n},m^2_{D_l},m^2_{L_o}) \nonumber
\end{eqnarray}
\begin{eqnarray}
B_{SRL}^{IJKL} &=& - {e^2\over 2s_W^2} Y_l^L Y_d^I \sum_{M=1}^3
\sum_{n=1}^2 K^{MI} K^{MJ\star} (Z_H^{1n})^2
D_2(m^2_{u_M},m^2_{H^+_n},M^2_W,0)\\
&-& {e\over 2s_W} Y_l^L \sum_{N=1}^3 \sum_{m,n=1}^2
\sum_{l=1}^6 Z_-^{2m} Z_+^{1n} Z_{\tilde \nu}^{LN}
Z_{\tilde \nu}^{KN\star} V_{dUC}^{RIlm} V_{dUC}^{LJln\star}
D_2(m^2_{C_m},m^2_{C_n},m^2_{U_l},m^2_{{\tilde \nu}_N})\nonumber\\
&-&\frac{1}{2} \sum_{m,n=1}^4\sum_{l,o=1}^6 (V_{lLN}^{LKon}
V_{lLN}^{RLom\star} +V_{lLN}^{LKom} V_{lLN}^{RLon\star})
V_{dDN}^{RIlm} V_{dDN}^{LJln\star}
D_2(m^2_{N_m},m^2_{N_n},m^2_{D_l},m^2_{L_o})\nonumber
\end{eqnarray}
\begin{eqnarray}
B_{TL}^{IJKL} &=& -{e\over 8 s_W} Y_l^L \sum_{N=1}^3 \sum_{m,n=1}^2
\sum_{l=1}^6 Z_+^{1n}  Z_-^{2m}  Z_{\tilde
\nu}^{LN} Z_{\tilde \nu}^{KN\star} V_{dUC}^{LIlm} V_{dUC}^{RJln\star}
m_{C_m} m_{C_n} D_0(m^2_{C_m},m^2_{C_n},m^2_{U_l},m^2_{{\tilde
\nu}_N})\nonumber\\
&-&\frac{1}{8} \sum_{m,n=1}^4\sum_{l,o=1}^6 (V_{lLN}^{LKon}
V_{lLN}^{RLom\star} -V_{lLN}^{LKom} V_{lLN}^{RLon\star})
V_{dDN}^{LIlm} V_{dDN}^{RJln\star} m_{N_m} m_{N_n}
D_0(m^2_{N_m},m^2_{N_n},m^2_{D_l},m^2_{L_o})\nonumber\\
\end{eqnarray}
\begin{eqnarray}
B_{TR}^{IJKL} &=& -{e\over 8 s_W} Y_l^K \sum_{N=1}^3 \sum_{m,n=1}^2
\sum_{l=1}^6 Z_+^{1m\star} Z_-^{2n\star} Z_{\tilde
\nu}^{LN} Z_{\tilde \nu}^{KN\star} V_{dUC}^{RIlm} V_{dUC}^{LJln\star}
m_{C_m} m_{C_n} D_0(m^2_{C_m},m^2_{C_n},m^2_{U_l},m^2_{{\tilde
\nu}_N}) \nonumber \\
&-&\frac{1}{8} \sum_{m,n=1}^4\sum_{l,o=1}^6 (V_{lLN}^{RKon}
V_{lLN}^{LLom\star} -V_{lLN}^{RKom} V_{lLN}^{LLon\star})
V_{dDN}^{RIlm} V_{dDN}^{LJln\star} m_{N_m} m_{N_n}
D_0(m^2_{N_m},m^2_{N_n},m^2_{D_l},m^2_{L_o}) \nonumber\\
\end{eqnarray}

\subsection{$Z$-penguins}

$ F_{ZX}$ are one-loop triangle-diagram contributions to the
$X$-handed ($X=L,R$) $\bar d^Id^JZ_\mu$ coupling.  The expression
below is valid only for the flavour violating case $I\neq J$ since,  
in order to simplify the formulae, we have dropped some terms appearing 
only for $I=J$.

\begin{eqnarray}
F_{ZL}^{IJ} &=& {e^3\over 4 s_W^3 c_W } \sum_{M=1}^3 K^{MI}
K^{MJ\star} \left[ \left(1-\frac{4s_W^2}{3}\right)
C_2(M_W^2,m_{u_M}^2,m_{u_M}^2) \right.\nonumber\\
&+& \left.\frac{8s_W^2}{3} m_{u_M}^2 C_0(M_W^2,m_{u_M}^2,m_{u_M}^2) +
6 c_W^2 C_2(M_W^2,M_W^2,m_{u_M}^2) \right]\nonumber\\
&+& {e^2 \sqrt{2} M_W\over c_W}\sum_{M=1}^3 K^{MI} K^{MJ\star}
Z_H^{22} Y_u^M m_{u_M} C_0(M_W^2,M_W^2,m_{u_M}^2) \nonumber\\
&-& {e\over 2 s_W c_W} \sum_{M=1}^3\sum_{l=1}^2 K^{MI} K^{MJ\star}
(Y_u^M)^2 (Z_H^{2l})^2 \left[ \frac{c_W^2-s_W^2}{2}
\left(C_2(m_{H^+_l}^2,m_{H^+_l}^2,m_{u_M}^2) + \frac{1}{2}\right)
\right.   \nonumber\\
&+& \left.   \frac{2 s_W^2}{3} \left(
C_2(m_{H^+_l}^2,m_{u_M}^2,m_{u_M}^2) - \frac{1}{2}\right) +
\left(1-\frac{4 s_W^2}{3}\right) m_{u_M}^2
C_0(m_{H^+_l}^2,m_{u_M}^2,m_{u_M}^2) \right] \nonumber\\
&+& \frac{1}{2} \sum_{l=1}^6\sum_{m,n=1}^2 V_{dUC}^{LIlm}
V_{dUC}^{LJln\star} \left[V_{ZCC}^{Lmn} \left(
C_2(m_{U_l}^2,m_{C_m}^2,m_{C_n}^2) -
\frac{1}{2}\right)\right.   \nonumber\\
&-&\left.   2 m_{C_m} m_{C_n} V_{ZCC}^{Rmn}
C_0(m_{U_l}^2,m_{C_m}^2,m_{C_n}^2)\right] \nonumber\\
&-&\frac{1}{2}\sum_{l,n=1}^6\sum_{m=1}^2 V_{ZUU}^{nl} V_{dUC}^{LIlm}
V_{dUC}^{LJnm\star} \left(C_2(m_{C_m}^2,m_{U_l}^2,m_{U_n}^2) +
\frac{1}{2}\right)\nonumber\\
&+& \frac{1}{2} \sum_{l=1}^6\sum_{m,n=1}^4 V_{dDN}^{LIlm}
V_{dDN}^{LJln\star} \left[ V_{ZNN}^{Lnm\star} \left(
C_2(m_{D_l}^2,m_{N_m}^2,m_{N_n}^2) - \frac{1}{2} \right)
\right.\nonumber\\
&+&\left.   2 V_{ZNN}^{Lnm} m_{N_m} m_{N_n}
C_0(m_{D_l}^2,m_{N_m}^2,m_{N_n}^2)\right] \nonumber\\
&-& \frac{1}{2} \sum_{l,n=1}^6\sum_{m=1}^4 V_{ZDD}^{ln} V_{dDN}^{LIlm}
V_{dDN}^{LJnm\star} \left(C_2(m_{N_m}^2,m_{D_l}^2,m_{D_n}^2) +
\frac{1}{2}\right) \nonumber\\
&+& \frac{2g_s^2}{3} \sum_{l,n=1}^6 V_{ZDD}^{ln} Z_D^{Il}
Z_D^{Jn\star} \left(C_2(m_G^2,m_{D_l}^2,m_{D_n}^2)+\frac{1}{2}\right)
\end{eqnarray}
\begin{eqnarray}
F_{ZR}^{IJ} &=& {e\over 4 s_W c_W} Y_d^I Y_d^J \sum_{M=1}^3
\sum_{l=1}^2 K^{MI} K^{MJ\star} (Z_H^{1l})^2 \left[ \left(1-\frac{4
s_W^2}{3}\right) C_2(m_{H^+_l}^2,m_{u_M}^2,m_{u_M}^2)\right.
\nonumber\\
&+&\left.   \frac{2 s_W^2}{3} m_{u_M}^2
C_0(m_{H^+_l}^2,m_{u_M}^2,m_{u_M}^2) - (c_W^2-s_W^2)
C_2(m_{H^+_l}^2,m_{H^+_l}^2,m_{u_M}^2) \right] \nonumber\\
&+& \frac{1}{2} \sum_{m,n=1}^2 \sum_{l=1}^6 V_{dUC}^{RIlm}
V_{dUC}^{RJln\star} \left[V_{ZCC}^{Rmn} \left(
C_2(m_{U_l}^2,m_{C_m}^2,m_{C_n}^2) - \frac{1}{2}\right)
\right.   \nonumber\\
&-&\left.    2 m_{C_m} m_{C_n} V_{ZCC}^{Lmn}
C_0(m_{U_l}^2,m_{C_m}^2,m_{C_n}^2)\right] \nonumber\\
&-&\frac{1}{2} \sum_{m=1}^2 \sum_{l,n=1}^6 V_{ZUU}^{nl} V_{dUC}^{RIlm}
V_{dUC}^{RJnm\star} \left( C_2(m_{C_m}^2,m_{U_l}^2,m_{U_n}^2) +
\frac{1}{2}\right) \nonumber\\
&-&\frac{1}{2} \sum_{m,n=1}^4 \sum_{l=1}^6 V_{dDN}^{RIlm}
V_{dDN}^{RJln\star} \left[ V_{ZNN}^{Lnm} \left(
C_2(m_{D_l}^2,m_{N_m}^2,m_{N_n}^2) - \frac{1}{2} \right)
\right.   \nonumber\\
&+& \left.   2 V_{ZNN}^{Lnm\star} m_{N_m} m_{N_n}
C_0(m_{D_l}^2,m_{N_m}^2,m_{N_n}^2)\right] \nonumber\\
&-&\frac{1}{2} \sum_{m=1}^4 \sum_{l,n=1}^6 V_{ZDD}^{ln} V_{dDN}^{RIlm}
V_{dDN}^{RJnm\star} C_2(m_{N_m}^2,m_{D_l}^2,m_{D_n}^2)\nonumber\\
&+&\frac{2 g_s^2}{3} \sum_{l,n=1}^6 V_{ZDD}^{ln} Z_D^{(I+3)l}
Z_D^{(J+3)n\star} \left(C_2(m_G^2,m_{D_l}^2,m_{D_n}^2) +
\frac{1}{2}\right)
\end{eqnarray}

\subsection{Higgs penguins}

$F_{HX}$ and $F_{AX}$ denote the $CP$-even and $CP$-odd
one-loop triangle-diagram contributions to the $X$-handed ($X=L,R$)
couplings $\bar d^Id^J H^0_k$ and $\bar d^Id^J A^0_k$ ($H_0^1\equiv H^0, H_0^2\equiv h^0, A_0^1\equiv A^0, A_0^2\equiv G^0$).  Appropriate
expressions are listed below -- please note that the explicit factor of
``$i$'' in the $CP$-odd higgs form factors is superficial and comes
from the definition of vertices in \ref{sec:feyrul}.  For the
$CP$-odd Higgs, the relevant vertices defined in this way are, for
real Lagrangian parameters, purely imaginary so that $iV$ is a real
number.

\begin{eqnarray}
F_{HL}^{IJk} &=& {e^2\over \sqrt{2} s_W^2} m_{d_I} \sum_{L=1}^3 K^{LJ}
K^{LI\star} \left( Z_R^{2k} m_{u_L} Y_u^L C_{12}(m_{u_L}^2,M_W^2) + {e^2
\over \sqrt{2} s_W^2} C_R^k C_{11}(M_W^2,m_{u_L}^2)\right) \nonumber\\
&-& {e^2\over 2\sqrt{2}s_W^2} Y_d^I \sum_{L=1}^3 \sum_{m=1}^2 A_M^{km}
Z_H^{1m} K^{LJ} K^{LI\star} C_2(m_{u_L}^2,M_W^2,m_{H^+_m}^2) \nonumber\\
&-& \frac{Z_R^{2k} Y_d^I}{\sqrt{2}} \sum_{L=1}^3 \sum_{m=1}^2 Z_H^{1m}
Z_H^{2m} (Y_u^L)^2 K^{LJ} K^{LI\star} \left(
C_2(m_{u_L}^2,m_{u_L}^2,m_{H^+_m}^2) + m_{u_L}^2
C_0(m_{u_L}^2,m_{u_L}^2,m_{H^+_m}^2)\right) \nonumber\\
&-& Y_d^I \sum_{L=1}^3 \sum_{m,n=1}^2 Z_H^{1n} Z_H^{2m}
V_{H^0H^+H^-}^{knm} Y_u^L m_{u_L} K^{LJ} K^{LI\star}
C_0(m_{u_L}^2,m_{H^+_m}^2,m_{H^+_n}^2) \nonumber\\
&-& \sum_{l,m=1}^2 \sum_{n=1}^6 V_{dUC}^{RInl\star} V_{dUC}^{LJnm}
\left( V_{SCC}^{mlk\star} C_2(m_{C_l}^2,m_{C_m}^2,m_{U_n}^2)
+ V_{SCC}^{lmk} m_{C_l} m_{C_m} C_0(m_{C_l}^2,m_{C_m}^2,m_{U_n}^2)
\right) \nonumber\\
&+& \sum_{n=1}^2 \sum_{l,m=1}^6 V_{SUU}^{klm} V_{dUC}^{RImn\star}
V_{dUC}^{LJln} m_{C_n} C_0(m_{U_l}^2,m_{U_m}^2,m_{C_n}^2) \nonumber\\
&-& \sum_{m,n=1}^4 \sum_{l=1}^6 V_{dDN}^{RInl\star} V_{dDN}^{LJnm}
\left( V_{SNN}^{lmk\star} C_2(m_{N_l}^2,m_{N_m}^2,m_{D_n}^2)
+ V_{SNN}^{lmk} m_{N_l} m_{N_m} C_0(m_{N_l}^2,m_{N_m}^2,m_{D_n}^2)
\right) \nonumber\\
&+& \sum_{n=1}^4 \sum_{l,m=1}^6 V_{SDD}^{klm} V_{dDN}^{RImn\star}
V_{dDN}^{LJln} m_{N_n} C_0(m_{D_l}^2,m_{D_m}^2,m_{N_n}^2) \nonumber\\
&-& \frac{8 g_s^2}{3} m_G \sum_{l,m=1}^6 V_{SDD}^{klm}
Z_D^{(I+3)m\star} Z_D^{Jl} C_0(m_{D_l}^2,m_{D_m}^2,m_G)
\end{eqnarray}
\begin{eqnarray}
F_{HR}^{IJk} &=& {e^2\over \sqrt{2} s_W^2} m_{d_J}
\sum_{L=1}^3 K^{LJ} K^{LI\star} \left(Z_R^{2k} m_{u_L} Y_u^L
C_{12}(m_{u_L}^2,M_W^2) + {e^2 \over \sqrt{2}s_W^2} C_R^k
C_{11}(M_W^2,m_{u_L}^2)\right) \nonumber\\
&+&{e^2\over 2\sqrt{2}s_W^2} Y_d^J \sum_{L=1}^3 \sum_{m=1}^2 A_M^{km}
Z_H^{1m} K^{LJ} K^{LI\star} C_2(m_{u_L}^2,M_W^2,m_{H^+_m}^2) \nonumber\\
&-& \frac{Z_R^{2k} Y_d^J}{\sqrt{2}} \sum_{L=1}^3 \sum_{m=1}^2 Z_H^{1m}
Z_H^{2m} (Y_u^L)^2 K^{LJ} K^{LI\star} \left(
C_2(m_{u_L}^2,m_{u_L}^2,m_{H^+_m}^2) + m_{u_L}^2
C_0(m_{u_L}^2,m_{u_L}^2,m_{H^+_m}^2)\right) \nonumber\\
&-& Y_d^J \sum_{L=1}^3 \sum_{m,n=1}^2 Z_H^{1m} Z_H^{2n}
V_{H^0H^+H^-}^{knm} Y_u^L m_{u_L} K^{LJ} K^{LI\star}
C_0(m_{u_L}^2,m_{H^+_m}^2,m_{H^+_n}^2)
\nonumber\\
&-& \sum_{l,m=1}^2 \sum_{n=1}^6 V_{dUC}^{LInl\star} V_{dUC}^{RJnm}
\left( V_{SCC}^{lmk} C_2(m_{C_l}^2,m_{C_m}^2,m_{U_n}^2) +
V_{SCC}^{mlk\star} m_{C_l} m_{C_m} C_0(m_{C_l}^2,m_{C_m}^2,m_{U_n}^2)
\right) \nonumber\\
&+& \sum_{n=1}^2 \sum_{l,m=1}^6 V_{SUU}^{klm} V_{dUC}^{LImn\star}
V_{dUC}^{RJln} m_{C_n} C_0(m_{U_l}^2,m_{U_m}^2,m_{C_n}^2) \nonumber\\
&-& \sum_{l,m=1}^4 \sum_{n=1}^6 V_{dDN}^{LInl\star} V_{dDN}^{RJnm}
\left(V_{SNN}^{lmk} C_2(m_{N_l}^2,m_{N_m}^2,m_{D_n}^2) +
V_{SNN}^{lmk\star} m_{N_l} m_{N_m} C_0(m_{N_l}^2,m_{N_m}^2,m_{D_n}^2)
\right) \nonumber\\
&+& \sum_{n=1}^4 \sum_{l,m=1}^6 V_{SDD}^{klm} V_{dDN}^{LImn\star}
V_{dDN}^{RJln} m_{N_n} C_0(m_{D_l}^2,m_{D_m}^2,m_{N_n}^2) \nonumber\\
&-& \frac{8 g_s^2}{3} m_G \sum_{l,m=1}^6 V_{SDD}^{klm} Z_D^{Im\star}
Z_D^{(J+3)l} C_0(m_{D_l}^2,m_{D_m}^2,m_G)
\end{eqnarray}
\begin{eqnarray}
F_{AL}^{IJk} &=& {e^2\over \sqrt{2}s_W^2} Z_H^{2k} m_{d_I}
\sum_{L=1}^3 m_{u_L} Y_u^L K^{LJ} K^{LI\star} C_0(m_{u_L}^2,m_{u_L}^2,M_W^2)
\nonumber\\
&-& {e \over 2s_W} Y_d^I \sum_{L=1}^3 K^{LJ} K^{LI\star} \left({e\over
\sqrt{2}s_W} Z_H^{1k} C_2(m_{u_L}^2,M_W^2,m_{H^+_k}^2) + Y_u^L m_{u_L}
M_W C_0(m_{u_L}^2,M_W^2,m_{H^+_1}^2) \right)
\nonumber\\
&-&\frac{Z_H^{2k} Y_d^I}{\sqrt{2}} \sum_{L=1}^3 \sum_{m=1}^2 Z_H^{1m}
Z_H^{2m} (Y_u^L)^2 K^{LJ} K^{LI\star} \left(
C_2(m_{u_L}^2,m_{u_L}^2,m_{H^+_m}^2) -m_{u_L}^2
C_0(m_{u_L}^2,m_{u_L}^2,m_{H^+_m}^2)\right) \nonumber\\
&-& i\sum_{l,m=1}^2 \sum_{n=1}^6 V_{dUC}^{RInl\star} V_{dUC}^{LJnm}
\left( V_{PCC}^{mlk\star} C_2(m_{C_l}^2,m_{C_m}^2,m_{U_n}^2)
- V_{PCC}^{lmk} m_{C_l} m_{C_m} C_0(m_{C_l}^2,m_{C_m}^2,m_{U_n}^2)
\right) \nonumber\\
&+& i\sum_{n=1}^2 \sum_{l,m=1}^6 V_{PUU}^{klm} V_{dUC}^{RImn\star}
V_{dUC}^{LJln} m_{C_n} C_0(m_{U_l}^2,m_{U_m}^2,m_{C_n}^2) \nonumber\\
&-&i \sum_{l,m=1}^4 \sum_{n=1}^6 V_{dDN}^{RInl\star} V_{dDN}^{LJnm}
\left(V_{PNN}^{lmk\star} C_2(m_{N_l}^2,m_{N_m}^2,m_{D_n}^2) -
V_{PNN}^{lmk} m_{N_l} m_{N_m} C_0(m_{N_l}^2,m_{N_m}^2,m_{D_n}^2)
\right) \nonumber\\
&+&i \sum_{n=1}^4 \sum_{l,m=1}^6 V_{PDD}^{klm} V_{dDN}^{RImn\star}
V_{dDN}^{LJln} m_{N_n} C_0(m_{D_l}^2,m_{D_m}^2,m_{N_n}^2) \nonumber\\
&-& {8ig_s^2\over 3} m_G \sum_{l,m=1}^6 V_{PDD}^{klm} Z_D^{(I+3)m\star}
Z_D^{Jl} C_0(m_{D_l}^2,m_{D_m}^2,m_G)
\end{eqnarray}
\begin{eqnarray}
F_{AR}^{IJk} &=& - {e^2\over \sqrt{2}s_W^2} Z_H^{2k} m_{d_J}
\sum_{L=1}^3 m_{u_L} Y_u^L K^{LJ} K^{LI\star} C_0(m_{u_L}^2,m_{u_L}^2,M_W^2)
\nonumber\\
&+& {e \over 2s_W} Y_d^J \sum_{L=1}^3 K^{LJ} K^{LI\star}
\left( {e\over \sqrt{2}s_W} Z_H^{1k}
C_2(m_{u_L}^2,M_W^2,m_{H^+_k}^2) + Y_u^L m_{u_L} M_W
C_0(m_{u_L}^2,M_W^2,m_{H^+_1}^2) \right) \nonumber\\
&+& \frac{Z_H^{2k} Y_d^J}{\sqrt{2}} \sum_{L=1}^3 \sum_{m=1}^2 Z_H^{1m}
Z_H^{2m} (Y_u^L)^2 K^{LJ} K^{LI\star} \left(
C_2(m_{u_L}^2,m_{u_L}^2,m_{H^+_m}^2) - m_{u_L}^2
C_0(m_{u_L}^2,m_{u_L}^2,m_{H^+_m}^2)\right) \nonumber\\
&+&i \sum_{l,m=1}^2 \sum_{n=1}^6 V_{dUC}^{LInl\star} V_{dUC}^{RJnm}
\left( V_{PCC}^{lmk} C_2(m_{C_l}^2,m_{C_m}^2,m_{U_n}^2) - 
V_{PCC}^{mlk\star} m_{C_l} m_{C_m} C_0(m_{C_l}^2,m_{C_m}^2,m_{U_n}^2)
\right) \nonumber\\
&+&i \sum_{n=1}^2 \sum_{l,m=1}^6 V_{PUU}^{klm} V_{dUC}^{LImn\star}
V_{dUC}^{RJln} m_{C_n} C_0(m_{U_l}^2,m_{U_m}^2,m_{C_n}^2) \nonumber\\
&+&i \sum_{l,m=1}^4 \sum_{n=1}^6 V_{dDN}^{LInl\star} V_{dDN}^{RJnm}
\left(V_{PNN}^{lmk} C_2(m_{N_l}^2,m_{N_m}^2,m_{D_n}^2) - 
V_{PNN}^{lmk\star} m_{N_l} m_{N_m} C_0(m_{N_l}^2,m_{N_m}^2,m_{D_n}^2)
\right) \nonumber\\
&+&i \sum_{n=1}^4 \sum_{l,m=1}^6 V_{PDD}^{klm} V_{dDN}^{LImn\star}
V_{dDN}^{RJln} m_{N_n} C_0(m_{D_l}^2,m_{D_m}^2,m_{N_n}^2) \nonumber\\
&-& {8ig_s^2\over 3} m_G \sum_{l,m=1}^6 V_{PDD}^{klm} Z_D^{Im\star}
Z_D^{(J+3)l} C_0(m_{D_l}^2,m_{D_m}^2,m_G)
\end{eqnarray}

\subsection{$d$ self-energy contributions}

Finally we list the formulae for the one-loop down quark self energy
contributions:
\begin{eqnarray}
\Sigma_{dV}^{IJ}&=&{e^2\over 2s_W^2} \sum_{L=1}^3 K^{LI} K^{LJ\star}
 B_1(0,m^2_{u_L},M^2_W)\nonumber\\
&+&\frac{1}{2} \sum_{L=1}^3 \sum_{k=1}^2 \left((Z_H^{2k})^2 (Y_u^L)^2
+Y_d^I Y_d^J (Z_H^{1k})^2\right) K^{LI} K^{LJ\star}
B_1(0,m^2_{u_L},m^2_{H^+_k}) \nonumber\\
&+&\frac{1}{2} \sum_{l=1}^2 \sum_{k=1}^6 (V_{dUC}^{LIkl}
V_{dUC}^{LJkl\star} +V_{dUC}^{RIkl} V_{dUC}^{RJkl\star})
B_1(0,m^2_{C_l},m^2_{U_k})\nonumber\\
&+&\frac{1}{2} \sum_{l=1}^4 \sum_{k=1}^6 (V_{dDN}^{LIkl}
V_{dDN}^{LJkl\star} +V_{dDN}^{RIkl} V_{dDN}^{RJkl\star})
B_1(0,m^2_{N_l},m^2_{D_k})\nonumber\\
&+&\frac{4g_s^2}{3} \sum_{k=1}^6 (Z_D^{Ik} Z_D^{Jk\star} +Z_D^{(I+3)k}
Z_D^{(J+3)k\star}) B_1(0,m^2_G,m^2_{D_k})
\end{eqnarray}
\begin{eqnarray}
\Sigma_{dA}^{IJ} &=& -{e^2\over 2s_W^2} \sum_{L=1}^3 K^{LI} K^{LJ\star} 
B_1(0,m^2_{u_L},M^2_W) \nonumber\\
&-&\frac{1}{2} \sum_{L=1}^3 \sum_{k=1}^2 \left( (Z_H^{2k})^2 (Y_u^L)^2
-Y_d^I Y_d^J (Z_H^{1k})^2\right) K^{LI} K^{LJ\star}
B_1(0,m^2_{u_L},m^2_{H^+_k}) \nonumber\\
&-&\frac{1}{2} \sum_{l=1}^2 \sum_{k=1}^6 (V_{dUC}^{LIkl}
V_{dUC}^{LJkl\star} -V_{dUC}^{RIkl} (V_{dUC}^{RJkl\star})
B_1(0,m^2_{C_l},m^2_{U_k})\nonumber\\
&-&\frac{1}{2} \sum_{l=1}^4 \sum_{k=1}^6 (V_{dDN}^{LIkl}
V_{dDN}^{LJkl\star} -V_{dDN}^{RIkl} V_{dDN}^{RJkl\star})
B_1(0,m^2_{N_l},m^2_{D_k})\nonumber\\
&-&\frac{4g_s^2}{3} \sum_{k=1}^6 (Z_D^{Ik} Z_D^{Jk\star} -Z_D^{(I+3)k}
Z_D^{(J+3)k\star}) B_1(0,m^2_G,m^2_{D_k})
\end{eqnarray}
\begin{eqnarray}
\Sigma_{dS}^{IJ}&=& \frac{1}{2} (Y_d^I+Y_d^J) \sum_{L=1}^3 \sum_{k=1}^2 
Z_H^{2k} Z_H^{1k} m_{u_L} Y_u^L K^{LI} K^{LJ\star}
B_0(0,m^2_{u_L},m^2_{H^+_k}) \nonumber\\
&-&\frac{1}{2} \sum_{l=1}^2 \sum_{k=1}^6 (V_{dUC}^{LIkl}
V_{dUC}^{RJkl\star} +V_{dUC}^{RIkl} V_{dUC}^{LJkl\star}) m_{C_l}
B_0(0,m^2_{C_l},m^2_{U_k})\nonumber\\
&-&\frac{1}{2} \sum_{l=1}^4 \sum_{k=1}^6 (V_{dDN}^{LIkl}
V_{dDN}^{RJkl\star} +V_{dDN}^{RIkl} V_{dDN}^{LJkl\star}) m_{N_l}
B_0(0,m^2_{N_l},m^2_{D_k})\nonumber\\
&+&\frac{4g_s^2}{3} m_G \sum_{k=1}^6 (Z_D^{Ik} Z_D^{(J+3)k\star}
+Z_D^{(I+3)k} Z_D^{Jk\star}) B_0(0,m^2_G,m^2_{D_k})
\end{eqnarray}
\begin{eqnarray}
\Sigma_{dP}^{IJ}&=&\frac{1}{2} (Y_d^I-Y_d^J) \sum_{L=1}^3 \sum_{k=1}^2 
Z_H^{2k} Z_H^{1k} m_{u_L} Y_u^L K^{LI} K^{LJ\star}
B_0(0,m^2_{u_L},m^2_{H^+_k}) \nonumber\\
&+&\frac{1}{2} \sum_{l=1}^2 \sum_{k=1}^6 (V_{dUC}^{LIkl}
V_{dUC}^{RJkl\star} - V_{dUC}^{RIkl} V_{dUC}^{LJkl\star}) m_{C_l}
B_0(0,m^2_{C_l},m^2_{U_k})\nonumber\\
&+&\frac{1}{2} \sum_{l=1}^4 \sum_{k=1}^6 (V_{dDN}^{LIkl}
V_{dDN}^{RJkl\star} - V_{dDN}^{RIkl} V_{dDN}^{LJkl\star}) m_{N_l}
B_0(0,m^2_{N_l},m^2_{D_k})\nonumber\\
&-&\frac{4g_s^2}{3} m_G \sum_{k=1}^6 (Z_D^{Ik} Z_D^{(J+3)k\star} -
Z_D^{(I+3)k} Z_D^{Jk\star}) B_0(0,m^2_G,m^2_{D_k})
\end{eqnarray}
%


\end{document}